\documentclass[aps,pre,twocolumn,groupedaddress,superscriptaddress, showpacs]{revtex4-1}

\usepackage{graphicx}
\usepackage{ textcomp }
\usepackage[usenames,dvipsnames]{color}

\usepackage{transparent}

\usepackage{tipa}

\usepackage{amsmath}

\usepackage{amsfonts}

\usepackage{amssymb}

\usepackage{dcolumn}

\usepackage[english]{babel}

\usepackage{array}

\usepackage{dsfont}

\begin{document}

\title{Self-similar space-filling sphere packings in three and four dimensions}

\author{D. V. St\"ager}

  \email{staegerd@ethz.ch}

  \affiliation{Computational Physics for Engineering Materials, IfB, ETH Zurich, Wolfgang-Pauli-Strasse 27, CH-8093 Zurich, Switzerland}

\author{H. J. Herrmann}

  \email{hans@ifb.baug.ethz.ch}

  \affiliation{Computational Physics for Engineering Materials, IfB, ETH Zurich, Wolfgang-Pauli-Strasse 27, CH-8093 Zurich, Switzerland}
  
  \affiliation{Departamento de F\'isica, Universidade Federal do Cear\'a, 60451-970 Fortaleza, Cear\'a, Brazil}

\begin{abstract}
Inversive geometry can be used to generate exactly self-similar space-filling sphere packings. We present a construction method in two dimensions and generalize it to search for packings in higher dimensions. We newly discover 29 three-dimensional and 13 four-dimensional topologies of which 10 and 5, respectively, are bearings. To distinguish and characterize the packing topologies, we numerically estimate their fractal dimensions and we analyze their contact networks.
\vspace{5mm}

\end{abstract}

\maketitle

\section{Introduction}

Space-filling sphere packings are idealized dense granular packings. They consist of hard spheres which, due to the specific spatial arrangement, leave no porosity in the limit of infinitesimally small spheres. Even though such perfect packings are difficult to realize experimentally, their size distribution and spatial arrangement serve as idealized references for dense packings of nearly spherical particles. Dense packings are needed in various industrial applications, such that they have been subject of many experimental and theoretical studies \cite{Ayer1965,Jodrey1985,
Yu1988,Ouchiyama1989,Soppe1990,Konakawa1990,Standish1991,
Yu1993,Anishchik1995,Elliott2002,Sobolev2010,Rahmani2014,Martin2014,Martin2015}.

Space-filling packings of spheres can be constructed in any dimension larger or equal to two. The simplest and most studied is the Apollonian Gasket \cite{Manna1991,Manna1991a,Borkovec1994,Anishchik1995,Sevier1995,
Doye2005,Varrato2011,Kranz2015}, where the basic idea is to iteratively fill each pore of a packing with the largest possible sphere. Some special packings, bipartite packings, have drawn attention due to their mechanical functionality. By definition, they only contain even loops of touching elements and thus, they allow for the rotation of all elements without any sliding friction in two \cite{Herrmann1990} and three dimensions \cite{Baram2004}. Therefore, they were called ``bearings'' and proposed as idealized models for seismic gaps  \cite{Herrmann1990,Oron2000,Baram2004,Baram2005,Astrom2000,Astroma2012}, regions between two tectonic plates with unexpectedly low seismic activity \cite{McCann1979,Society1982}.

Here, we generate space-filling packings using inversive geometry which coercively leads to exactly self-similar fractal packings. Note that there are further methods to construct space-filling fractal packings such as those producing the Kleinian circle packings \cite{Sevier1995,Parker1995}, the random bipartite packings of Ref.\ \cite{Baram2005}, or osculatory packings \cite{Kausch1970,Pickover1989,Varrato2011}. Our work is motivated by previous studies on exactly self-similar space-filling bearings, as elaborated in the next paragraph.

First, two families, F1 and F2, of two-dimensional disk bearings were found \cite{Herrmann1990}, having smallest loops of size four. Later, these two families were generalized to smallest loops of any even size \cite{Oron2000}. In both Refs.\ \cite{Herrmann1990,Oron2000}, the packings are constructed in a strip geometry by iteratively applying conformal transformations, namely reflections, translations, and inversions, to some initially placed disks. In contrast, Ref.\ \cite{Borkovec1994} presents a way to construct a self-similar space-filling packing using inversive geometry only. In particular, they construct the 3D Apollonian Gasket which is based on the geometry of a tetrahedron. The packing is constructed inside a sphere, unlike the space-filling disk packings constructed on a strip in Refs.\ \cite{Herrmann1990,Oron2000}. Later, this construction technique was generalized to other Platonic solids than the tetrahedron \cite{Baram2004a}, leading to a total of five packings, of which one is a bearing. When applied in two dimensions, this approach turns out to generate the topologies of family F1 presented in Ref.\ \cite{Herrmann1990}, i.e., bipartite packings with smallest loops of size four. Note that any of these packings can be inverted as a whole to switch from the strip configuration to a packing enclosed by a circle and vice versa.

For this work, we carefully studied both of the families F1 and F2 presented in Ref.\ \cite{Herrmann1990} for smallest loop size four and for arbitrary even smallest loop size in Ref.\ \cite{Oron2000}, to find a general method to construct all of them enclosed by a circle and using inversive geometry only, inspired by the construction methods in Refs.\ \cite{Borkovec1994,Baram2004a}. In comparison to the construction on a strip, where different configurations generally differ in the length of the periodic unit cell, the construction inside a circle leads to configurations which are based on different regular polygons. This method can be straightforwardly extended to any higher dimension and we used it to find further packings in three and four dimensions. We care about four-dimensional packings because by cutting them with a three dimensional hyperplane, a three dimensional packing can be obtained. In that way, any 4D packing serves as a source for further 3D packings.

The order of content in this paper is the following. In Sec.\ \ref{sec:basic_idea}, we give a basic idea of how we use inversive geometry to generate space-filling packings. In Sec.\ \ref{sec:inversion}, we provide the necessary knowledge about inversive geometry that is needed to understand our work. In Sec.\ \ref{sec:constraints_on_setup}, we show the constraints which our construction method needs to fulfill to lead to a space-filing packing. In Sec.\ \ref{sec:method}, we describe how to construct all two-dimensional packings of both families F1 and F2 enclosed by a circle and generalize our method to higher dimensions. In Sec.\ \ref{sec:characterization}, we present the newly discovered packings and characterize them by estimating their fractal dimensions and analyzing their contact networks. Futhermore, we discuss in Sec.\ \ref{sec:modify_packings} how one can generate further variations of packings. At last, we give final remarks in Sec.\ \ref{sec:conclusion}.

\section{basic idea of generating a packing}\label{sec:basic_idea}

\begin{figure}[]
\begin{center}
	\includegraphics[width=\columnwidth]{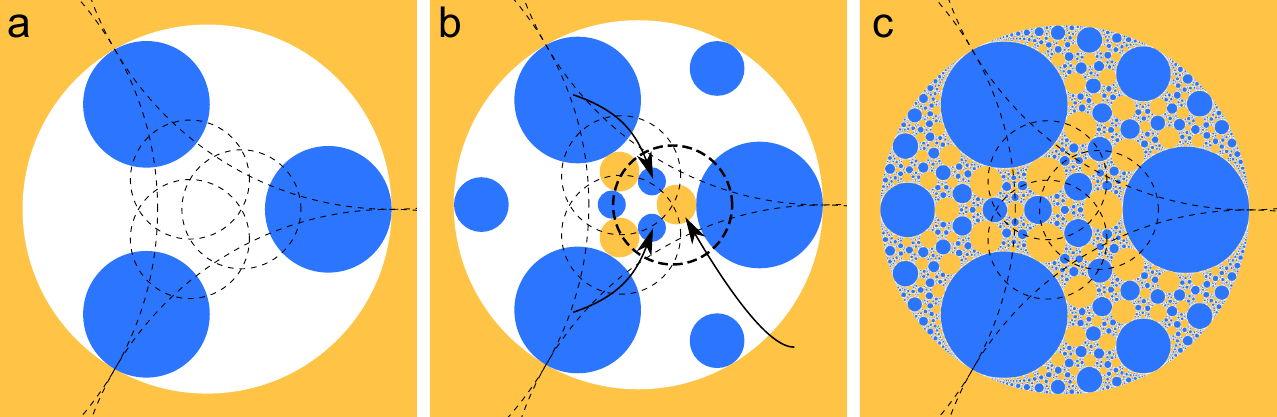}
\end{center}
\caption{
\label{fig:basic_idea} Basic idea of generating a space-filling packing. (a) Generating setup: unit circle hole and initially placed disks (filled) as seeds together with inversion circles (dashed). (b) New disks are generated by inverting the seeds at the inversion circles. Arrows lead from seeds to their inversions with respect to the highlighted inversion circle. (c) Space-filling packing as a result from infinite iterative inversions of newly generated disks.
}
\end{figure}

In two dimensions, each packing is constructed inside the unit circle, which can be seen as a hole that we aim to fill. We start by placing initial disks inside it. These disks and the unit circle hole itself act as seeds of the packing, out of which new disks will be generated by inversion. For that, a group of inversion circles is used. They form, together with the seeds, the generating setup of the packing as shown in Fig.\ \ref{fig:basic_idea}a. The seeds are inverted at the inversion circles to generate new disks as shown in Fig.\ \ref{fig:basic_idea}b. The newly generated disks can again be inverted to obtain further disks. This can be repeated infinitely many times till all space is filled as shown in Fig.\ \ref{fig:basic_idea}c. Along the same line, one can generate space-filling packings in any higher dimension.

To end up with a space-filling packing, the generating setup needs to fulfill certain constraints on how disks and inversion circles are placed. Before we discuss these constraints in Sec.\ \ref{sec:constraints_on_setup}, we will provide the necessary understanding of inversive geometry in the following section.

\section{circle inversion}\label{sec:inversion}

We will explain important properties of circle inversion considering a single inversion circle in Sec.\ \ref{sec:inversion_basic} and multiple inversion circles in Sec.\ \ref{sec:inversion_multiple}. Note that all properties explained in the following hold analogously for sphere inversions in 3D and in any higher dimension.

\subsection{Basic properties}\label{sec:inversion_basic}

\begin{figure*}[]
\begin{center}
	\includegraphics[width=2\columnwidth]{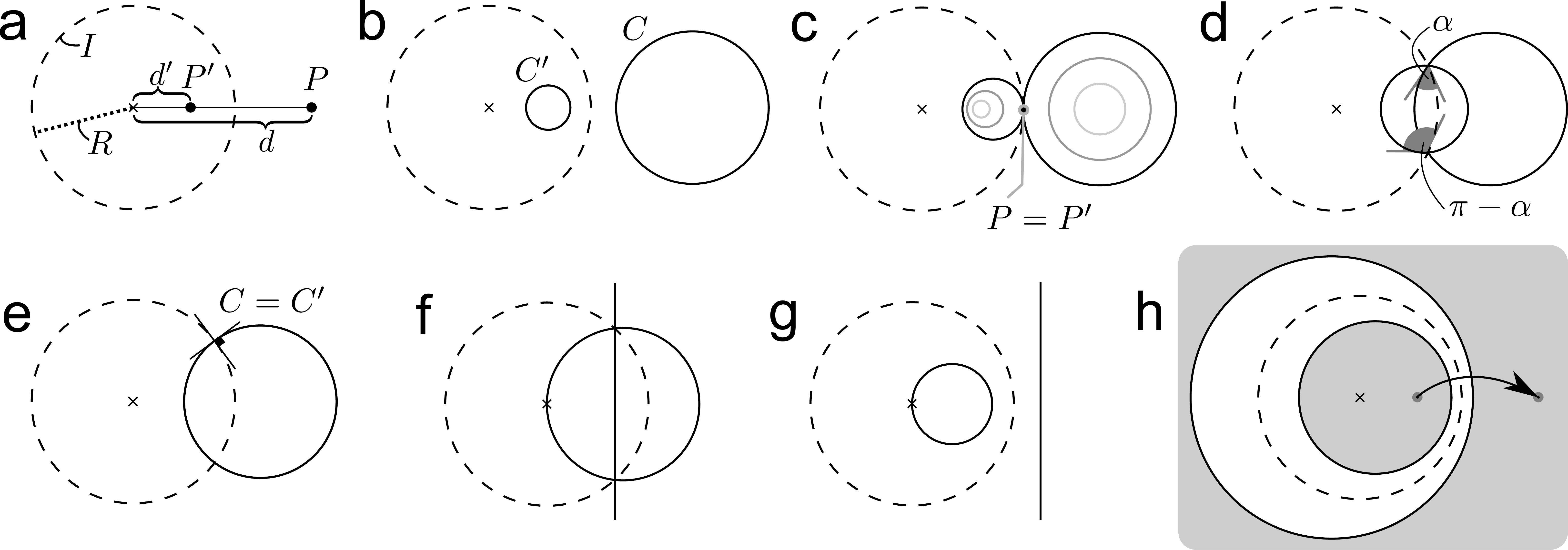}
\end{center}
\caption{
\label{fig:inversion_examples} Examples of circle inversion. (a) Inversion of a single point $P$ at an inversion circle $I$ (dashed). The distance to the center of $I$ is $d$ for point $P$ and $d'=R^2/d$ for its inverse $P'$, where $R$ is the radius of $I$. (b) A circle $C$ laying outside $I$ leads to the inverse circle $C'$ laying inside and vice versa. (c) $C$ touching $I$ from the outside results in $C'$ touching $I$ from the inside. (d) $C$ intersecting $I$ with angle $\alpha$ leads to $C'$ intersecting $I$ with angle $\pi-\alpha$. (e) If $C$ is perpendicular to $I$, $C'$ is identical to $C$. (f,g) $C$ touching the center of $I$ results in $C'$ being a line, i.e., a circle with infinite radius. (h) If a circle includes the center of $I$, the inversion turns the circle inside out, i.e., the area inside the circle is mapped onto the area outside its inverse, as indicated by the arrow.
}
\end{figure*}

Figure\ \ref{fig:inversion_examples} shows the basic properties of circle inversion. If we invert a single point $P$ at an inversion circle $I$ as shown in Fig.\ \ref{fig:inversion_examples}a, the image $P'$ will lay in the same direction as $P$ when looking from the center of $I$. But its distance to the center will be $d'=R^2/d$, where $d$ is the distance of $P$ from the center of $I$ and $R$ is its radius. Therefore, if $P$ lays outside $I$, then $P'$ lays inside, and vice versa. Thus, an inversion circle $I$ divides the space into two sections that are mapped onto each other, namely the inside and the outside of $I$.

Circle inversion is a conformal mapping. The inverse of a circle $C$ with respect to an inversion circle $I$ is a circle $C'$, as shown in Fig.\ \ref{fig:inversion_examples}b. Since inversion is a self-inverse mapping, the inverse of $C'$ at $I$ is the original circle $C$. Mathematically, circle inversion is made simple by using a transformation of the positions and radii of the circles into so called inversive coordinates. Using inversive coordinates, one obtains the inverse circle $C'$ by multiplying a matrix, that depends on the inversion circle $I$, with the original circle $C$ (details in Appendix \ref{app:math_of_inversion}).

Figure \ref{fig:inversion_examples}c shows, that if a circle $C$ touches $I$ from the outside, $C'$ will touch $I$ in the same point from the inside. Note that the inverse $P'$ of any point $P$ that lays on $I$ is identical to $P$. Furthermore, the center of $C$ is in general not mapped onto the center of $C'$, as indicated by the concentric grey circles and their inverses in Fig.\ \ref{fig:inversion_examples}c.

If $C$ intersects $I$ with an angle $\alpha$, $C'$ will intersect $I$ at the same points with an angle $\pi-\alpha$ (Fig.\ \ref{fig:inversion_examples}d). Therefore, if $C$ is perpendicular to $I$, $C'$ is identical to $C$ (Fig.\ \ref{fig:inversion_examples}e).

The center of $I$ is inverted to the point at infinity. Therefore, if $C$ touches the center of $I$, it is mapped onto a line, i.e., onto a circle with infinite radius as shown in Figs.\ \ref{fig:inversion_examples}f and \ref{fig:inversion_examples}g.

Finally, we treat every circle either as a disk or a hole, referring to the area inside or outside the circle, respectively. To distinguish between disks and holes mathematically, we assign a positive radius $r>0$ to disks and a negative radius $r<0$ to holes, such that the surface of both disks and holes is the circle with radius $|r|$. This convention is meaningful, since if the center of an inversion circle $I$ lays inside a disk with radius $r>0$, the inversion (mathematical details in Appendix \ref{app:math_of_inversion}) turns the disk inside out into a hole, i.e., the area inside the surface of the disk is mapped onto the area outside the surface of the hole (Fig.\ \ref{fig:inversion_examples}h).

\subsection{Multiple inversion circles}\label{sec:inversion_multiple}
A single inversion circle divides space into two sections that are mapped onto each other, the inside and the outside. The situation is more complex if multiple inversion circles are present. Let us deal with two inversion circles $I_1$ and $I_2$.

\begin{figure}[]
\begin{center}
	\includegraphics[width=\columnwidth]{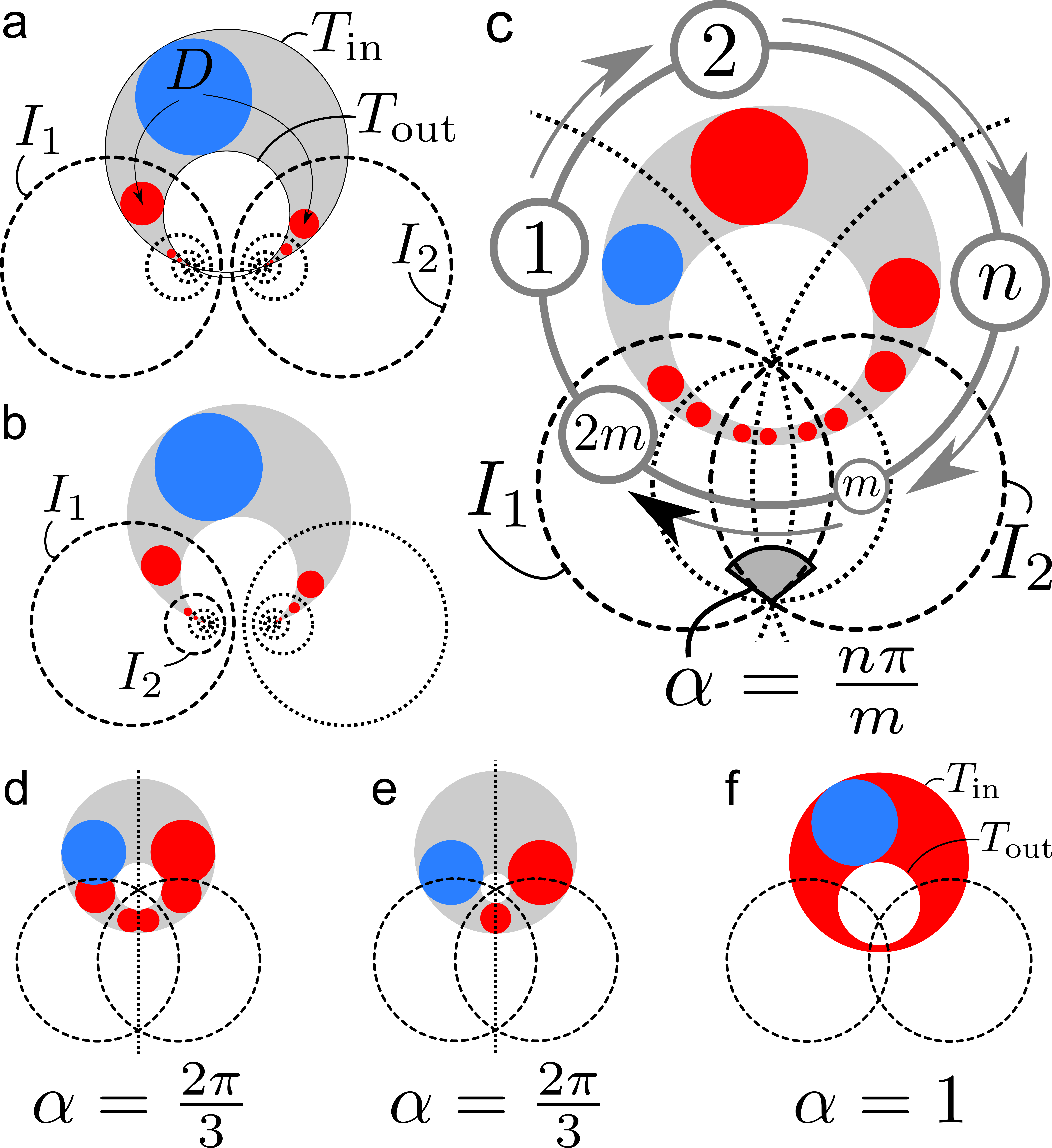}
\end{center}
\caption{
\label{fig:two_inversion_circles_examples} Examples of two inversion circles $I_1$ and $I_2$ (dashed). (a) If $I_1$ and $I_2$ do not overlap, they divide space into an infinity of non-overlapping sections, where the inversion circles are section borders themselves. The other section borders (dotted) are obtained by iteratively inverting $I_1$ and $I_2$ at one another. A disk $D$ placed in the area both outside $I_1$ and $I_2$, which is a single section, leads to infinitely many non overlapping disks, one in each section. All disks lay outside $T_{out}$ and inside $T_{in}$, which are the two circles tangent to $D$ and perpendicular to both $I_1$ and $I_2$. (b) If an inversion circle lays inside another, the space inside and outside both of them is divided into infinite sections. (c) If $I_1$ and $I_2$ intersect with an angle $\alpha = n \pi/ m$, where $n$ and $m$ are integers without common prime factors and $1 \leq n < m$, space is divided into $2m$ sections. $n$ of these sections lay outside of both $I_1$ and $I_2$. A disk placed inside one of the sections only, will result in a total of $2m$ disks, one in each section. (d) If a disk is placed on a section border but not perpendicular to it, overlapping disks will be generated. (e) A disk placed perpendicular to a section border, will result in a total of $m$ disks. (f) $\alpha=1$: If $\alpha/\pi$ is not a rational number, space is divided into infinite infinitesimally small sections. Thus, any initially placed disk results in completely filling the space which is both inside the circle $T_{in}$ and outside the circle $T_{out}$ with infinitely many partially overlapping disks.
}
\end{figure}

If $I_1$ and $I_2$ do not overlap, any disk laying outside of both of them can be iteratively inverted at $I_1$ and $I_2$ an infinite number of times leading to infinitely many new disks as shown in Fig.\ \ref{fig:two_inversion_circles_examples}a. Like the initial disk itself, all its images will lay both inside the circle $T_{in}$ and outside the circle $T_{out}$ which are the only two circles that are tangent to the initial disk and perpendicular to both $I_1$ and $I_2$. We can see that $I_1$ and $I_2$ together divide space into an infinity of non-overlapping areas, which we will call ``sections'', that are mapped onto each other. $I_1$ and $I_2$ themselves are section borders. Besides, the other section borders that divide space are obtained by iteratively inverting $I_1$ and $I_2$ at one another (dotted circles in Fig.\ \ref{fig:two_inversion_circles_examples}a). Each of these section borders lays inside $I_1$ or $I_2$, such that the space outside both of them remains a single section. As shown in Fig.\ \ref{fig:two_inversion_circles_examples}b, in the case where $I_2$ lays inside $I_1$ or vice versa, space is also divided into infinite sections, but some of the section borders lay outside of both inversion circles.

If $I_1$ and $I_2$ intersect, the intersecting angle $\alpha$ determines in how many sections space is divided. If $\alpha = n \pi/ m$, where $n$ and $m$ are integers without common prime factors and $1 \leq n < m$, space is divided into $2m$ sections as shown in Fig.\ \ref{fig:two_inversion_circles_examples}c. $n$ of these sections lay outside both $I_1$ and $I_2$, as well as in their overlapping region. $m-n$ sections lay exclusively in $I_1$, and another $m-n$ sections exclusively in $I_2$. A disk that is placed inside one of the sections will by iterative inversions at $I_1$ and $I_2$ lead to a disk in every section. If a disk is placed on a section border but not perpendicular to it, overlapping disks will be generated (Fig.\ \ref{fig:two_inversion_circles_examples}d). To avoid overlapping, one should place disks only inside sections or, as shown in Fig.\ \ref{fig:two_inversion_circles_examples}e, perpendicular to section borders. If $\alpha/\pi$ is not a rational number, space is divided into infinite infinitesimally small sections, such that any initially placed disk would lead to infinitely many partially overlapping images as shown in Fig.\ \ref{fig:two_inversion_circles_examples}f for $\alpha=1$.

\section{Constraints on generating setup}\label{sec:constraints_on_setup}

As we have seen in Fig.\ \ref{fig:basic_idea}, a generating setup consists of seeds and inversion circles. To lead to a non-overlapping space-filling packing, the setup needs to fulfill two constraints, as we explain in the following two paragraphs.

\begin{figure}[]
\begin{center}
	\includegraphics[width=\columnwidth]{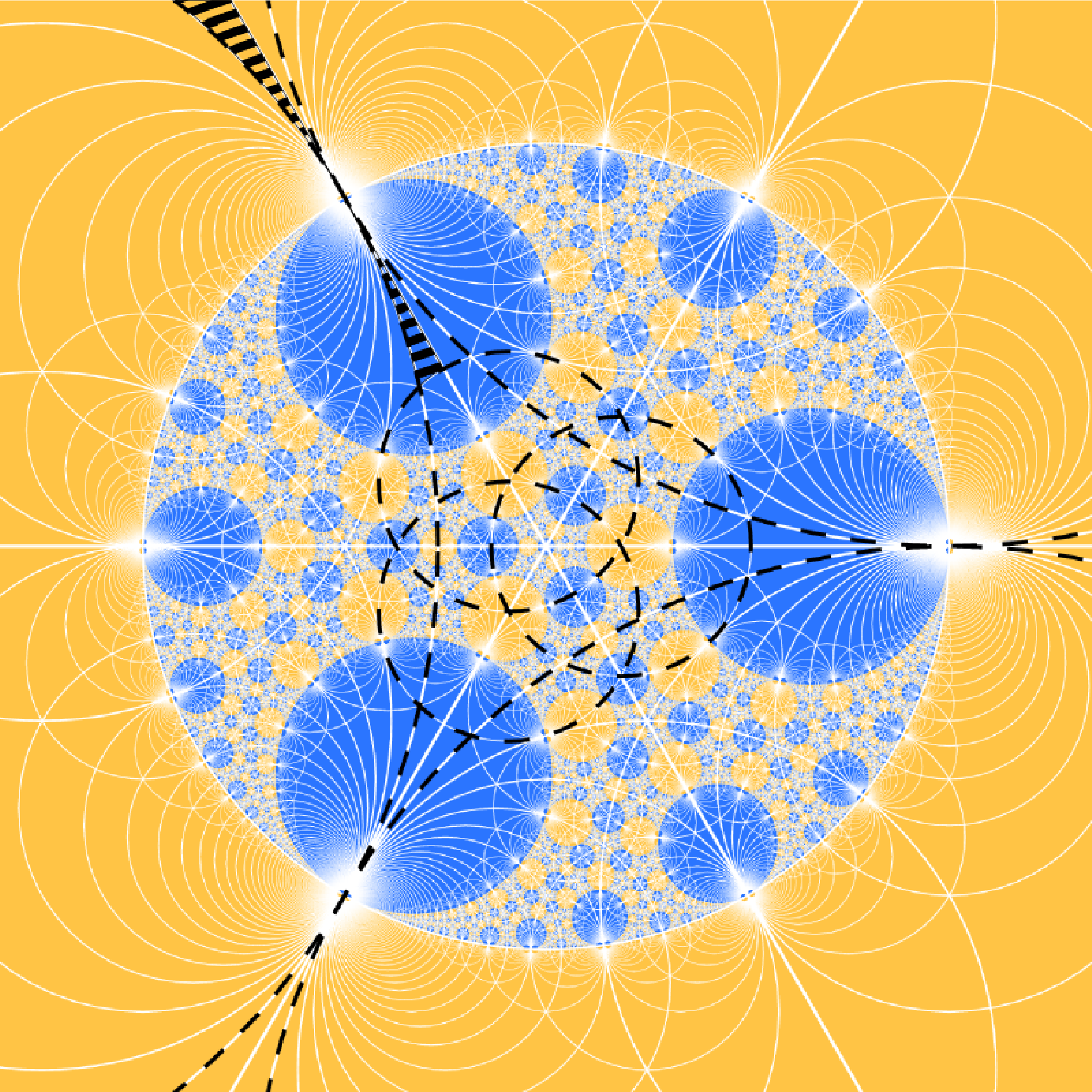}
\end{center}
\caption{
\label{fig:constraints_setup_coverage} The inversion circles (dashed) of the generating setup divide space into sections (section borders with radius $r>0.01$ shown in white). The sections form groups, here two, colored in blue (dark grey) and yellow (light grey), where sections of the same group are mapped onto each other by a certain sequence of inversions. Each group contains at least one section (striped area) which is outside all inversion circles. Therefore, covering the space outside all inversion circles guarantees all sections to be eventually covered, leading to a space-filling packing.
}
\end{figure}

First, to guarantee that the resulting packing is space-filling, the seeds and inversion circles together need to cover all space. This is a conjecture from previous studies \cite{Baram2004,Baram2005a}, which we will prove here with a detailed explanation. We have seen in Fig.\ \ref{fig:two_inversion_circles_examples} how multiple inversion circles divide space into sections. Figure \ref{fig:constraints_setup_coverage} shows a generating setup with all sections created by its inversion circles. One can group these sections such that each section of a certain group can be mapped onto any other section of the same group by a sequence of inversions. In Fig.\ \ref{fig:constraints_setup_coverage}, we find two groups. If we cover a single section of a certain group, the whole group will eventually be covered by iterative inversions. In every generating setup, every group contains at least one section that is outside all inversion circles. This follows from the fact that any section that is inside an inversion circle, can be inverted at this inversion circle to be mapped onto a larger section outside of it. If this inverse of the section lays again inside any of the inversion circles, we can invert it again to map it outside of that inversion circle onto an even larger section. This can be repeated till we end up with a section that lays outside all inversion circles. Therefore if we cover all space outside all inversion circles with seeds, all space will eventually be covered by iterative inversions.

\begin{figure}[]
\begin{center}
	\includegraphics[width=\columnwidth]{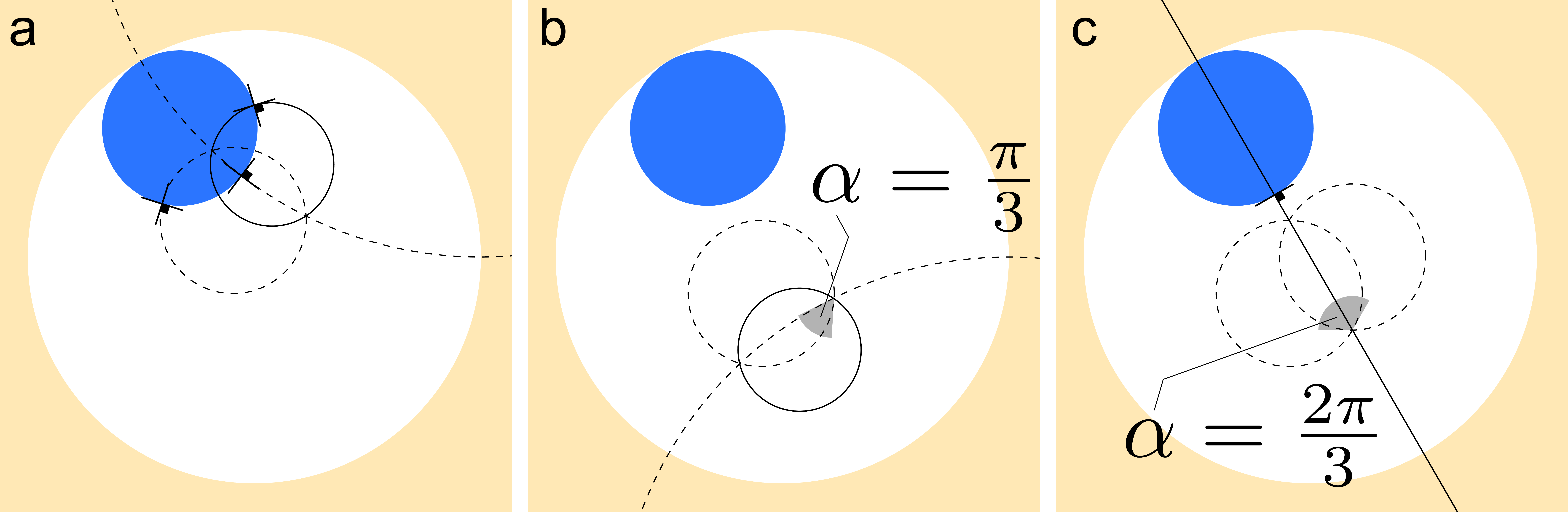}
\end{center}
\caption{
\label{fig:constraints_setup_angles} Constraints on intersecting angles in the generating setup. (a) A seed (filled) is always placed perpendicular to some inversion circles (dashed) and therefore will be perpendicular to all section borders created by these inversion circles, since they are a mapping of the inversion circles themselves. (b) For any two intersecting inversion circles, we allow in general only intersecting angles of $\alpha = \pi/m$ with $m\geq 2$, such that the area outside both inversion circles remains a single section. (c) For intersecting inversion circles with a mirror symmetry of the generating setup between them, we allow intersecting angles $\alpha = 2\pi/m$ with any integer $m\geq3$, such that even in the case of odd $m$, the section border outside both inversion circles is a mirror line of the setup and is therefore by default perpendicular to all intersecting seeds.
}
\end{figure}

Second, we need to avoid the partial overlap of disks (compare Fig.\ \ref{fig:two_inversion_circles_examples}d). Partial overlap is avoided if every placed seed is perpendicular to all intersecting section borders created by the inversion circles as shown in Fig.\ \ref{fig:constraints_setup_coverage}. We achieve this by the following strategy. As shown in Fig.\ \ref{fig:two_inversion_circles_examples}c, if two inversion circles intersect with an angle $\alpha = n \pi/ m$, where $n$ and $m$ are integers without common prime factors and $1 \leq n < m$, space is divided into $2m$ sections. Importantly, $n$ of these sections are outside both inversion circles. Because of that, we generally only allow intersecting angles $\alpha = \pi/m$ with $m\geq 2$. This ensures that the area outside both inversion circles remains a single section. Given that, we place seeds perpendicular to some inversion circles and at the same time on the outside of all others. This ensures two things. On one hand, as indicated in Fig.\ \ref{fig:constraints_setup_angles}a, the seed will be perpendicular to all the section borders created by the inversion circles that it is perpendicular to, since these borders are a mapping of the inversion circles themselves. On the other hand, the combination of all the inversion circles that are outside of the seed will never create a section border that will intersect the seed as indicated in Fig.\ \ref{fig:constraints_setup_angles}b. In addition to that, in the special case where by construction we have a mirror symmetry between two intersecting inversion circles, we allow intersecting angles of $\alpha = 2 \pi/ m$ with any integer $m \geq 3$. For even $m$, the area outside both inversion circles remains a single section. And for odd $m$, the area outside both inversion circles is divided into just two sections by a line along a mirror symmetry of the setup, which by default is perpendicular to all intersecting seeds, as shown in Fig.\ \ref{fig:constraints_setup_angles}c.

\section{How to construct generating setups}\label{sec:method}
We first describe how we construct generating setups for all two-dimensional packings of families F1 and F2 in the Sec.\ \ref{sec:defining_description_2D} before we generalize to higher dimensions in Sec.\ \ref{sec:generalization_higher_dimensions}. How to obtain the exact positions and radii of elements is shown in detail in Appendix \ref{app:solve_for_spatial_details}.

\subsection{Construction of 2D generating setups}\label{sec:defining_description_2D}

\begin{figure}[]
\begin{center}
	\includegraphics[width=\columnwidth]{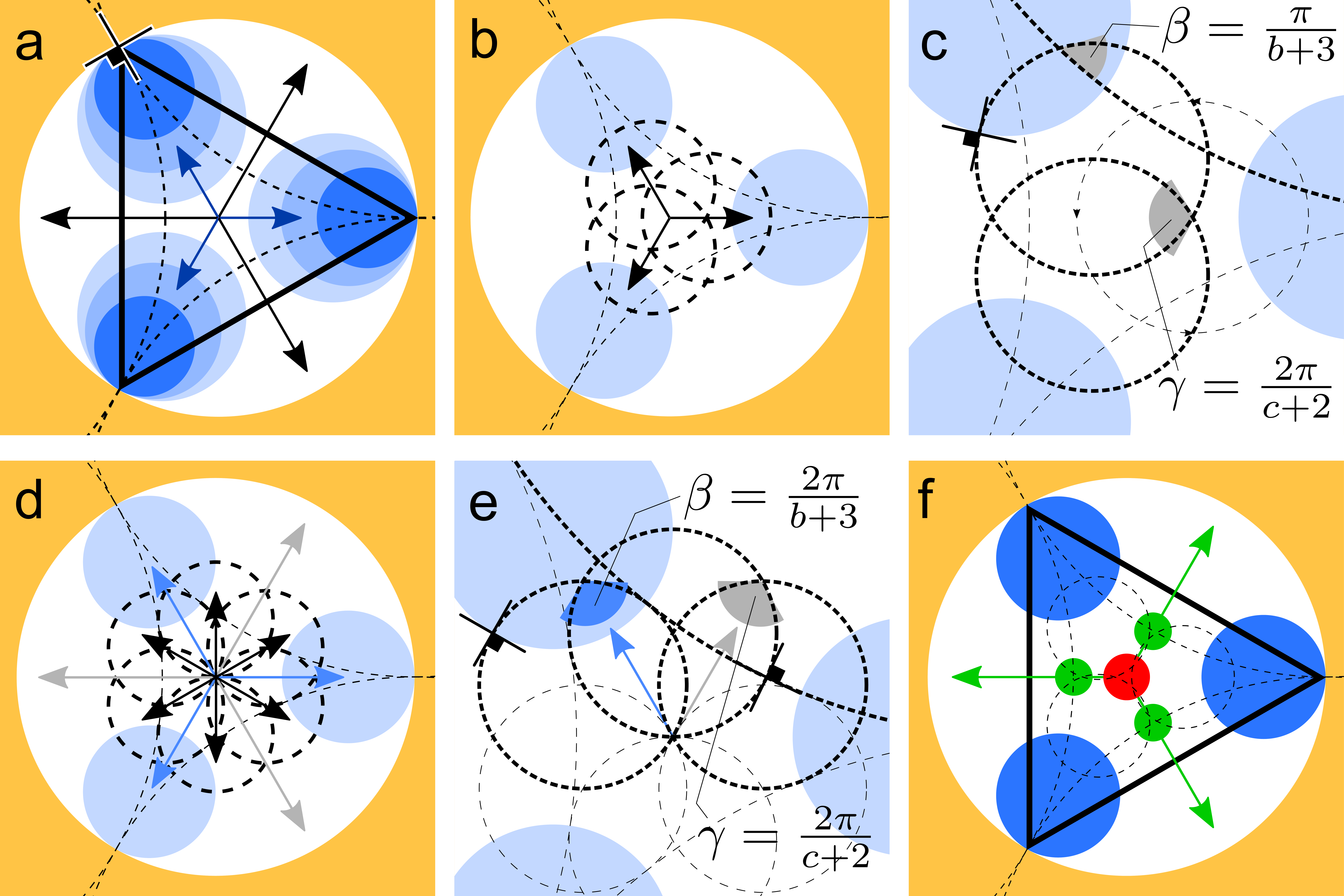}
\end{center}
\caption{
\label{fig:2D_circular_packings1} Construction of 2D generating setups. (a) Every setup is based on a regular polygon. The primary seeds are tangent to the unit circle and lay in the direction of the vertices of the polygon. The outer inversion circles lay in the direction of its edges. For F1, the inner inversion circles lay in the direction of the primary seeds (b) with intersecting angles as indicated in (c). For F2, the inner inversion circles lay in directions between the primary seeds and the outer inversion circles (d) with intersecting angles as indicated in (e). (f) Some packings need additional seeds, which can only lay in the direction of the edges of the polygon or in its center.
}
\end{figure}

Every setup is based on the geometry of a regular polygon, as the one shown in Fig.\ \ref{fig:2D_circular_packings1}a. In addition to the unit circle hole as a first seed, we always have another set of seeds tangent to it, which we call primary seeds (blue colored in all figures). They lay in the direction of the vertices of the polygon. Furthermore, there is a set of inversion circles which lay in the direction of the faces of the polygon, and which have their centers outside the unit circle, such that we call them the outer inversion circles. As shown in Fig.\ \ref{fig:2D_circular_packings1}a, the outer inversion circles are perpendicular to the unit circle and the nearest primary seeds, which already defines their size and positions. For a given regular polygon, they are identical for both packing families F1 and F2.

In addition to the outer inversion circles, there are the inner inversion circles, that lay completely inside the unit circle. The size of the primary seeds depends on the inner inversion circles. The inner inversion circles are different for the two families. 

For F1, as shown in Fig.\ \ref{fig:2D_circular_packings1}b, the inner inversion circles lay in the directions of the primary seeds. Figure \ref{fig:2D_circular_packings1}c indicates the allowed intersecting angles between different inversion circles. The inner inversion circles are perpendicular to the nearest primary seeds and intersect the outer inversion circles with angle $\beta=\pi/(3+b)$ with an integer $b\geq0$. Nearest neighboring inner inversion circles intersect each other with angle $\gamma=2\pi/(2+c)$ with an integer $c\geq0$. This never leads to overlapping disks since there is a mirror symmetry between any two nearest neighbors of the inner inversion circles (compare Fig.\ \ref{fig:constraints_setup_angles}c). In the special case of $c=0$, inner inversion circles that intersect each other with angle $\gamma=\pi$ are actually identical circles, i.e., they collapse to a single inversion circle.

For F2, as shown in Fig.\ \ref{fig:2D_circular_packings1}d, the inner inversion circles lay in directions between the primary seeds and the outer inversion circles. As shown in Fig.\ \ref{fig:2D_circular_packings1}e, they are always perpendicular to the outer inversion circles. Inner inversion circles nearest to a certain primary seed intersect each other with angle $\beta=2\pi/(3+b)$ with an integer $b\geq0$. The ones nearest to a certain outer inversion circle intersect each other with angle $\gamma=2\pi/(2+c)$ with an integer $c\geq0$, collapsing to a single circle for $c=0$.

\begin{figure}[]
\begin{center}
	\includegraphics[width=\columnwidth]{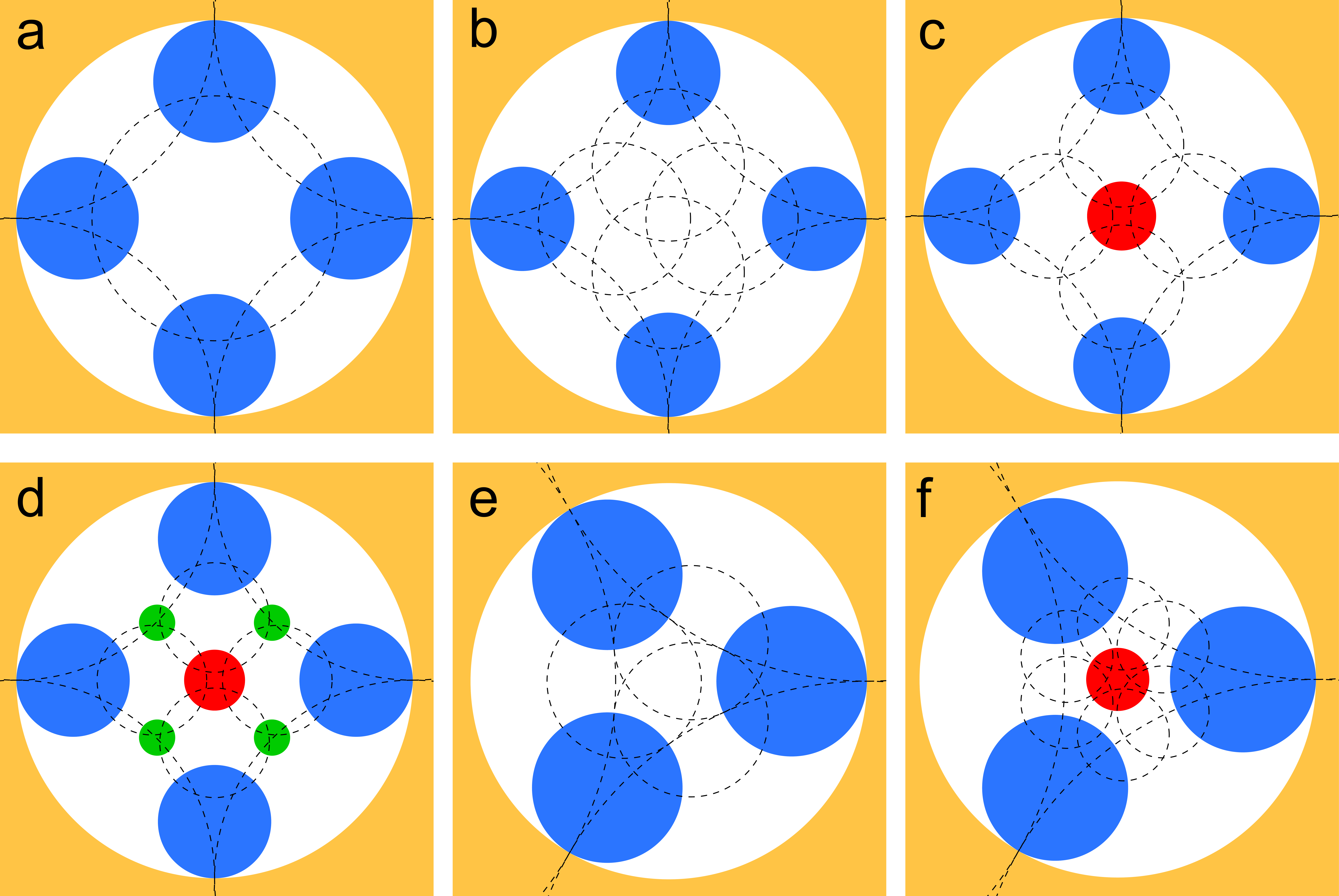}
\end{center}
\caption{
\label{fig:2D_circular_packings2} Different generating setups in the format (b,c) for F1 based on the square from (a) to (d): (0,0), (0,1), (0,3), (1,3), and for F2 based on the triangle in (e) (0,0) and (f) (1,1).
}
\end{figure}

The number of vertices $N$ of the polygon, the choice of family, and the parameters $b$ and $c$ together define the packing. Our integer parameters can be expressed by the ones used in Ref.\ \cite{Oron2000}, such that $N=3+n_1$, $b=n_2$, and $c=(l-4)/2$, where $l$ is the size of the smallest loops of the packing. For some parameters, the generating setup does not cover the whole space, but one can add additional seeds to still end up with a space-filling packing. These seeds need to be perpendicular to all inversion circles surrounding the uncovered area. Such additional seeds can only be needed in the center of the packing or in the direction of the edges of the polygon as shown in Fig.\ \ref{fig:2D_circular_packings1}f. Some more examples of generating setups are shown in Fig.\ \ref{fig:2D_circular_packings2}. Note that both increasing $b$ (compare Figs.\ \ref{fig:2D_circular_packings2}c and \ref{fig:2D_circular_packings2}d) and increasing $c$ (compare Figs.\ \ref{fig:2D_circular_packings2}b and \ref{fig:2D_circular_packings2}c) can lead to more seeds being necessary to cover all space.

\subsection{Generalization to higher dimensions}\label{sec:generalization_higher_dimensions}

In two dimensions, one can create an infinite number of distinct space-filling topologies for both families F1 and F2. In higher dimensions, we only find a finite number, because for some choices of basic shapes and parameters we are unable to generate a space-filling packing while avoiding overlapping spheres. Therefore the generalized method for higher dimensions only acts as a tool to search for possible generating setups.

In analogy to the choice of a regular convex polygon as base, one can choose a regular convex $n$-polytope when considering the $n$-dimensional space. In 3D, there are the Platonic Solids consisting of the tetrahedron, cube, octahedron, dodecahedron, and icosahedron. In 4D, we have the 5-cell, 8-cell, 16-cell, 24-cell, 120-cell, and 600-cell. In five and higher dimensions, there are only the three shapes that exist in any dimension $n$, i.e., the $n$-simplex (triangle (n=2), tetrahedron (n=3), 5-cell (n=4)), the $n$-cube (square (n=2), cube (n=3), 8-cell (n=4)), and the $n$-orthoplex (square (n=2), octahedron (n=3), 16-cell (n=4)). In 2D, we placed the outer inversion circles in the direction of the edges of the chosen polygon and the primary seeds in the direction of the vertices. In higher dimensions, we have more options to position the corresponding elements. In 3D, one can position them on either the vertices, edges, or faces. Note that some shapes are dual to each other, such as the cube and the octahedron in 3D. Therefore the positions of the faces of a cube are identical to the vertices of a octahedron, and vice versa. Placing the outer inversion spheres on the faces of a cube and the primary seeds on its vertices, is equal to placing the outer inversion spheres on the vertices of an octahedron and the primary seeds on its faces. Thus, to avoid finding each generating setup twice because of shape dualities, we consider every shape, but place the primary seeds always on lower dimensional elements than the outer inversion spheres. The vertices are the 0-dimensional elements, followed by one-dimensional edges, two-dimensional faces, etc. This way, in 3D and 4D, it happens that all possible generating setups have the primary seeds positioned at the vertices of the chosen shape, and the outer inversion spheres therefore at edges, faces, or cells (only for 4D).

\begin{figure}[]
\begin{center}
	\includegraphics[width=\columnwidth]{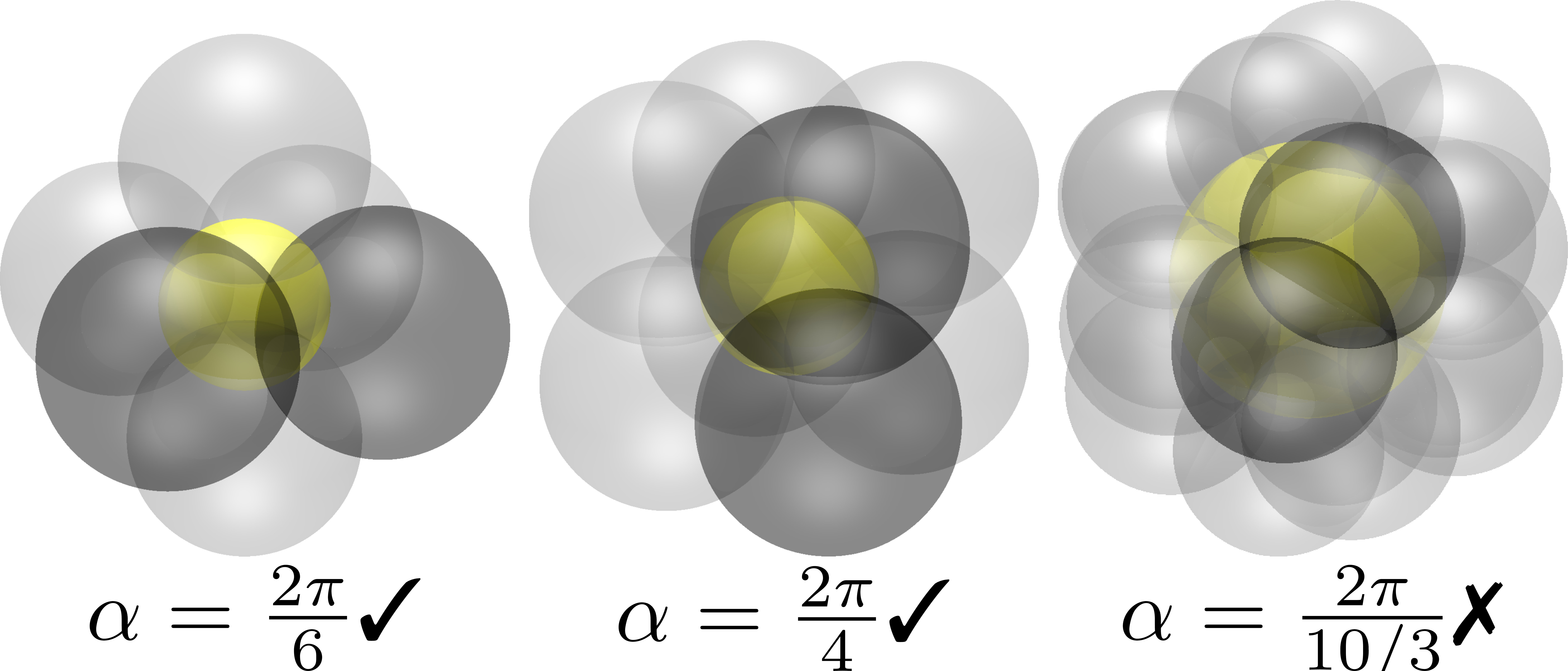}
\end{center}
\caption{
\label{fig:3D_outer_overlap} Outer inversion spheres are only allowed to intersect each other with angles $\alpha = 2\pi /m$ with $m\geq 3$ being an integer (check-mark). For non-integer $m$ (cross), no packing can be constructed. From left to right: Outer inversion spheres and unit sphere hole (central) of setups based on the cube, octahedron, and icosahedron, with outer inversion spheres at the faces of the shape and primary seeds at the vertices.
}
\end{figure}

In every generating setup in three or more dimensions, the outer inversion spheres will intersect each other. Therefore, in contrast to 2D, one additionally needs to check the intersecting angles of all intersecting outer inversion spheres. All intersecting angles $\alpha = 2 \pi / m$ with an integer $m \geq 3$ are allowed (compare Fig.\ \ref{fig:constraints_setup_angles}c), since we have a mirror symmetry by default between any pair of outer inversion spheres. Figure \ref{fig:3D_outer_overlap} shows examples of the intersecting angles of outer inversion spheres. If a forbidden angle exists, no generating setup can be derived.

\begin{figure*}[]
\begin{center}
	\includegraphics[width=2\columnwidth]{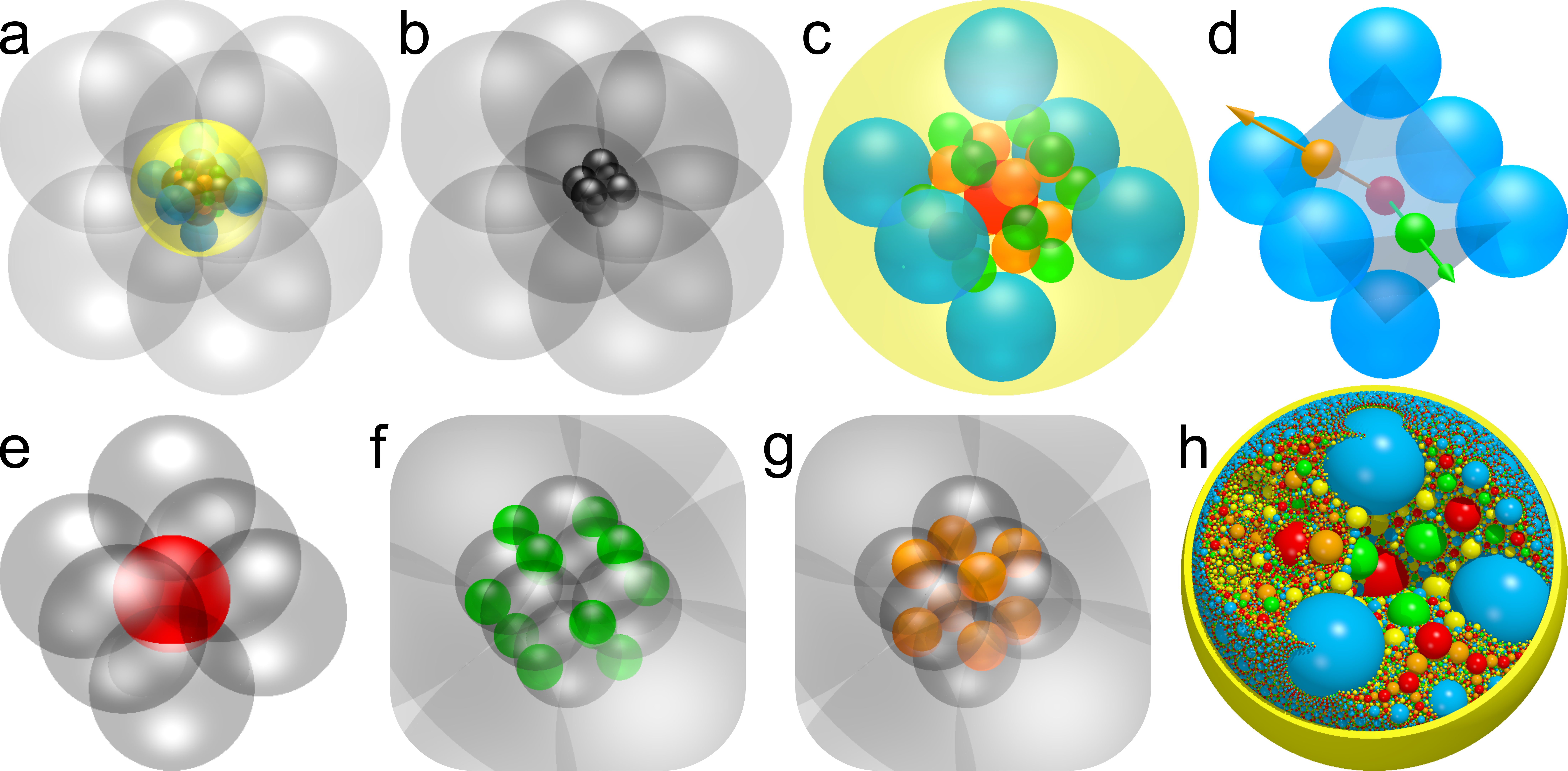}
\end{center}
\caption{
\label{fig:complete_setup} (a) Complete 3D generating setup based on the octahedron with outer inversion spheres at its faces with $b=0$ and $c=4$. (b) Inversion spheres only. (c) Seeds only. (d) The convex polytope that has the primary seeds as vertices defines possible positions for additional seeds. They can only be at the center or in the direction of the edges, faces, etc., of the polytope. (e) An additional seed in the center would need to be perpendicular to the inner inversion spheres. Other additional seeds in the direction of the edges (f) or faces (g) of the polytope in (d) need to be perpendicular to both inner and outer inversion spheres. (h) Resulting packing with lowest considered radius $r_{min}=0.005$. Unit sphere visualized as an open shell and some spheres removed to allow looking inside the packing.
}
\end{figure*}

After the choice of positions for the primary seeds and outer inversion spheres, one can for both F1 and F2, choose a combination of $b$ and $c$ what defines the inner inversion spheres, the primary seeds, and all additional seeds, such that the whole setup is defined. A complete 3D generating setup can be seen in Fig.\ \ref{fig:complete_setup}a, with Fig.\ \ref{fig:complete_setup}b showing only the inversion spheres and Fig.\ \ref{fig:complete_setup}c only the seeds. The possible positions for additional seeds, which are needed to cover potentially uncovered space, are defined by the positions of the primary seeds. For symmetric reasons, they can only lay in the center or in the directions of the edges, faces, etc., of the convex polytope that has the primary seeds as vertices, as indicated in Fig.\ \ref{fig:complete_setup}d. A central seed needs to be perpendicular to the inner inversion spheres (Fig.\ \ref{fig:complete_setup}e), whereas the other additional seeds need to be perpendicular to the inner and outer inversion spheres (Figs.\ \ref{fig:complete_setup}f and \ref{fig:complete_setup}g).

After calculating the sizes and positions of all elements in the generating setup, one needs to check if all the constraints, that ensure that a space-filling packing can be generated, are fulfilled: Seeds are not allowed to intersect, what in contrast to 2D can happen in higher dimensions. Furthermore, one needs to check all intersecting angles between inversion spheres, since in addition to the defined angles by $b$ and $c$, intersections of not nearest neighbors of inversion spheres with a forbidden angle might exist. If all these constraints are fulfilled, a space-filling packing can be generated as the one shown in Fig.\ \ref{fig:complete_setup}h.

The search for generating setups for any choice of positions of outer inversion spheres and primary seeds can be started with parameters $b=c=0$. If a valid setup is found, one can continue the search by increasing either $b$ or $c$. At some point, additional seeds are needed (compare Fig.\ \ref{fig:2D_circular_packings2}). While increasing $b$ or $c$ further, some seeds will finally overlap each other, and one knows that no further setups can be found by further increasing the parameters.

\section{Discovered Packings}\label{sec:characterization}

We searched for generating setups in three and four dimensions. Some generating setups turned out to give the exact same packings and some lead to different packings but identical topologies, i.e., equal fractal dimension and contact network properties. With our generalized method, we find 54 generating setups in 3D which lead to 34 distinct topologies, of which 29 are new discoveries. In 4D, we find 29 generating setups leading to 13 distinct topologies, none of them reported before.

We characterize the packings of all generating setups in different ways to show the topological differences. We determine the fractal dimension as described in Sec.\ \ref{sec:df}. In Sec.\ \ref{sec:clusters}, we analyze the contact network where we check for isolated spheres, count the number of connected clusters, find the smallest loops of each cluster, and check if clusters are bipartite. An overview of all found generating setups with all characterizations can be found in Table \ref{tab:3D} and \ref{tab:4D} for 3D and 4D, respectively.

\subsection{Fractal dimension}\label{sec:df}
The fractal dimension can be estimated from the total number of spheres $N(r)$, their cumulative surface $s(r)$, or the remaining porosity $p(r)$ of a packing of spheres with radius larger $r$. Details to generate packings down to a smallest radius computationally efficiently can be found in Appendix \ref{app:computation}. The unit sphere hole needs to be treated as a sphere of radius one with the corresponding surface but a negative volume, i.e., one needs to add its volume to the remaining porosity.

\begin{figure}[]
\begin{center}
	\includegraphics[width=\columnwidth]{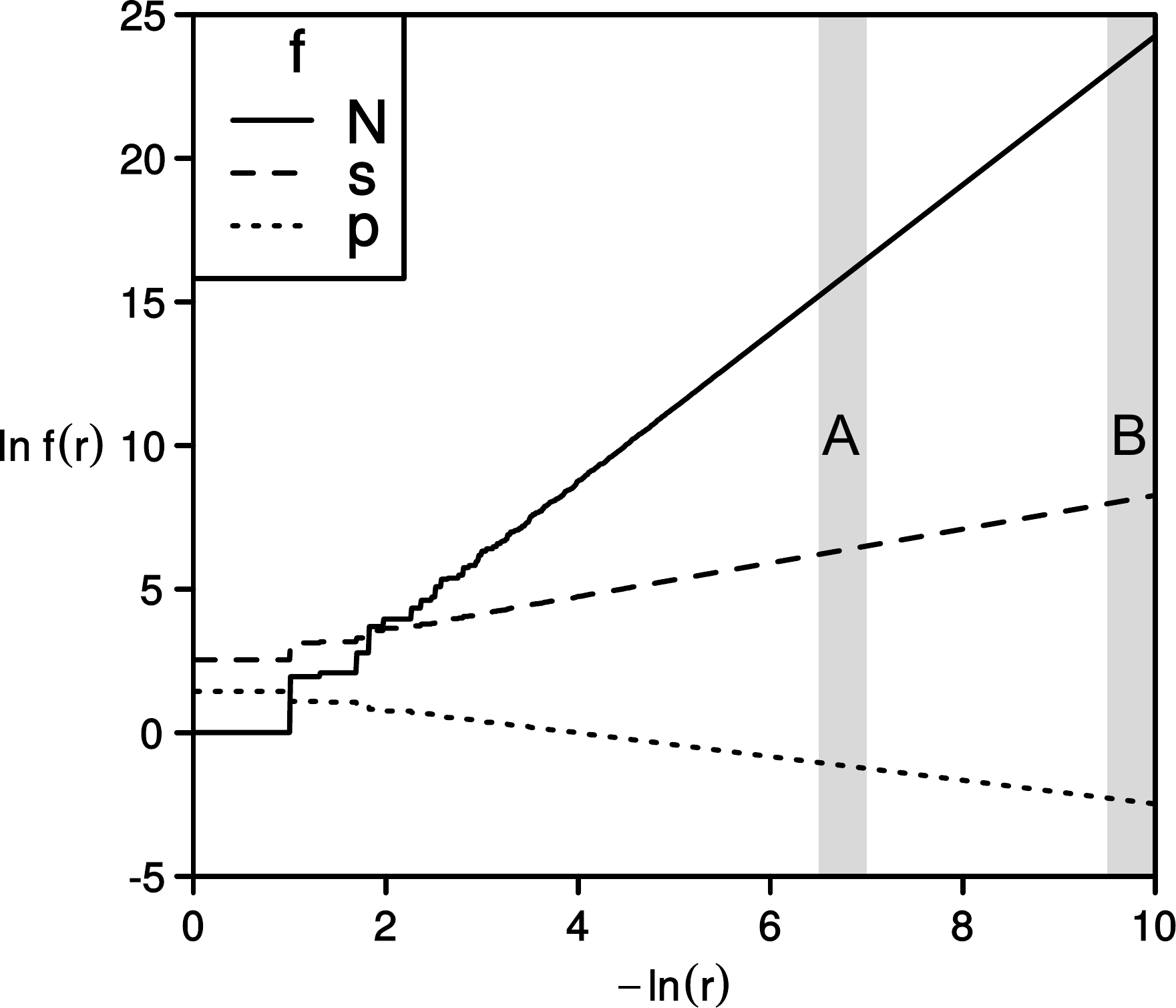}
\end{center}
\caption{
\label{fig:Nps} Total number of spheres $N$, cumulative surface $s$, and remaining porosity $p$ as a function of the lowest considered radius $r$ for the previously known 3D packing of Ref.\ \cite{Baram2004}. From the slope in a double logarithmic scale one can extract estimates for the fractal dimensions. Estimates from moving intervals from A to B can be found in Fig.\ \ref{fig:df_estimates}.
}
\end{figure}

\begin{figure}[]
\begin{center}
	\includegraphics[width=\columnwidth]{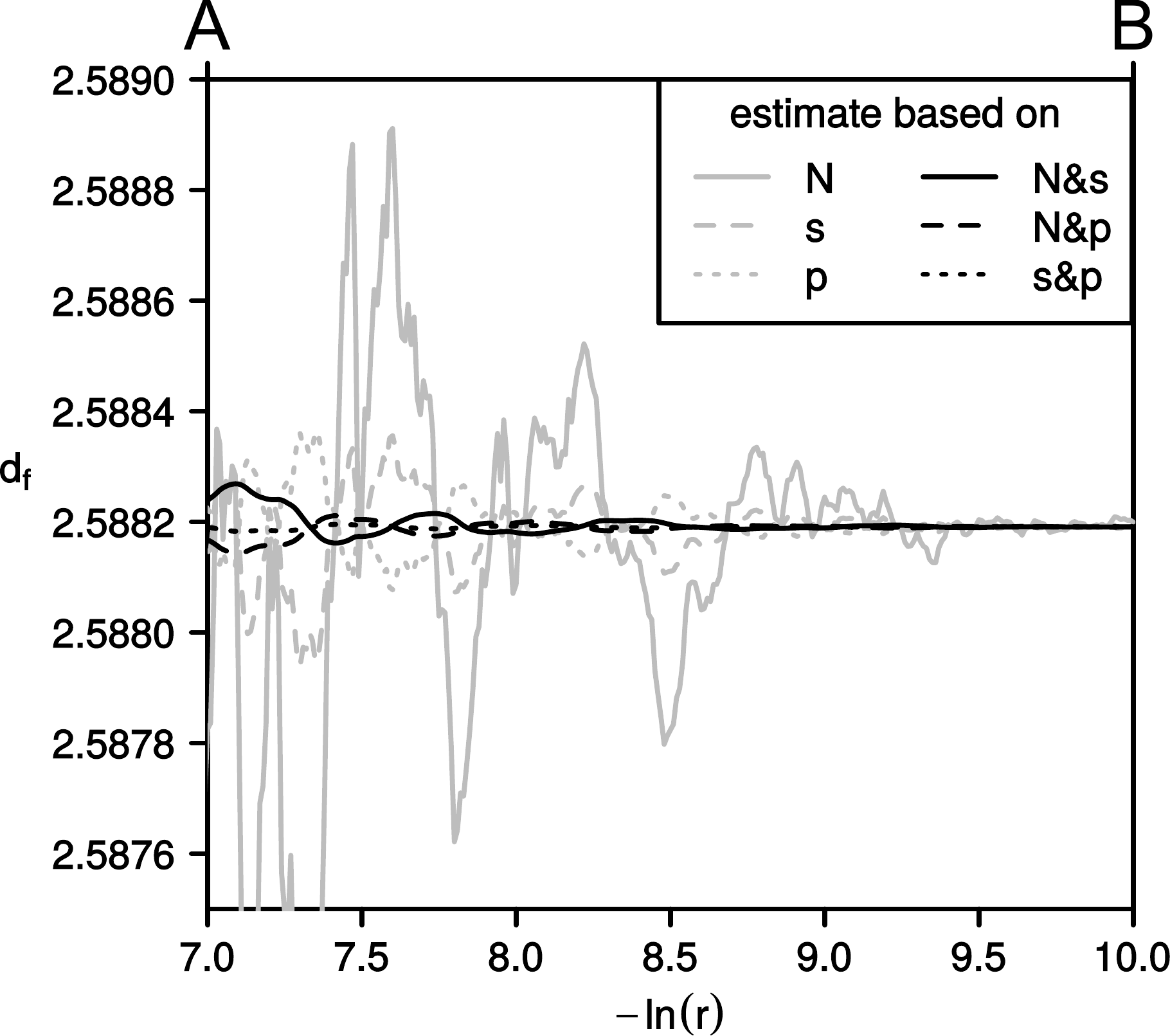}
\end{center}
\caption{
\label{fig:df_estimates} Estimates $\hat{d}_f$ of the fractal dimensions based on the single functions $N(r)$, $s(r)$, and $p(r)$, and combined estimates based on pairs of these functions. Estimates are extracted from moving intervals from A to B shown in Fig.\ \ref{fig:Nps}, for the previously known 3D packing of Ref.\ \cite{Baram2004}.
}
\end{figure}

The functions $N(r)$, $s(r)$, and $p(r)$ follow the asymptotic behaviors
\begin{equation}
\label{eq:asymptotic_behavior}
N(r) \sim r^{-d_f}, \ \ s(r) \sim r^{-d_f-D+1} , \ \ p(r) \sim r^{-d_f-D},
\end{equation}
where $d_f$ is the fractal dimension and $D$ are the dimensions of space. An estimate $\hat{d}_f$ for the fractal dimension can be obtained from the slope of these functions on a double logarithmic scale as the one shown in Fig.\ \ref{fig:Nps}. We extract estimates on the intervals $[r,r\sqrt{e}]$, which we move toward lower $r$ to see the fluctuations of $\hat{d}_f(r)$ to judge its accuracy, as shown in Fig.\ \ref{fig:df_estimates}. This approach is also used in Ref.\ \cite{Borkovec1994}. We further use a way to combine different estimates to improve accuracy. From the asymptotic behavior in Eq.\ (\ref{eq:asymptotic_behavior}), we assume that the errors $\Delta \hat{d}_{f_a}=\hat{d}_{f_a}-d_f$ and $\Delta \hat{d}_{f_b}=\hat{d}_{f_b}-d_f$ on two estimates $\hat{d}_{f_a}$ and $\hat{d}_{f_b}$ based on a certain interval of two different functions $a(r) \sim r^{A-d_f}$ and $b(r) \sim r^{B-d_f}$, respectively, relate as
\begin{equation}
\label{eq:df_error_relation}
\frac{\Delta \hat{d}_{f_a}}{A-d_f} = \frac{\Delta \hat{d}_{f_b}}{B-d_f}.
\end{equation}
We use Eq.\ (\ref{eq:df_error_relation}) to define a combined estimate $\hat{d}_{f_{a\&b}}$ based on both functions $a(r)$ and $b(r)$ as
\begin{equation}
\label{eq:df_better_guess}
\hat{d}_{f_{a\&b}}=x \hat{d}_{f_a}+(1-x) \hat{d}_{f_b}, \ \text{with} \ x=\frac{B-\hat{d}_{f_b}}{B-A+\hat{d}_{f_a}-\hat{d}_{f_b}},
\end{equation}
which mostly shows a smoother behavior, as shown in Fig.\ \ref{fig:df_estimates}.

\begin{figure}[]
\begin{center}
	\includegraphics[width=\columnwidth]{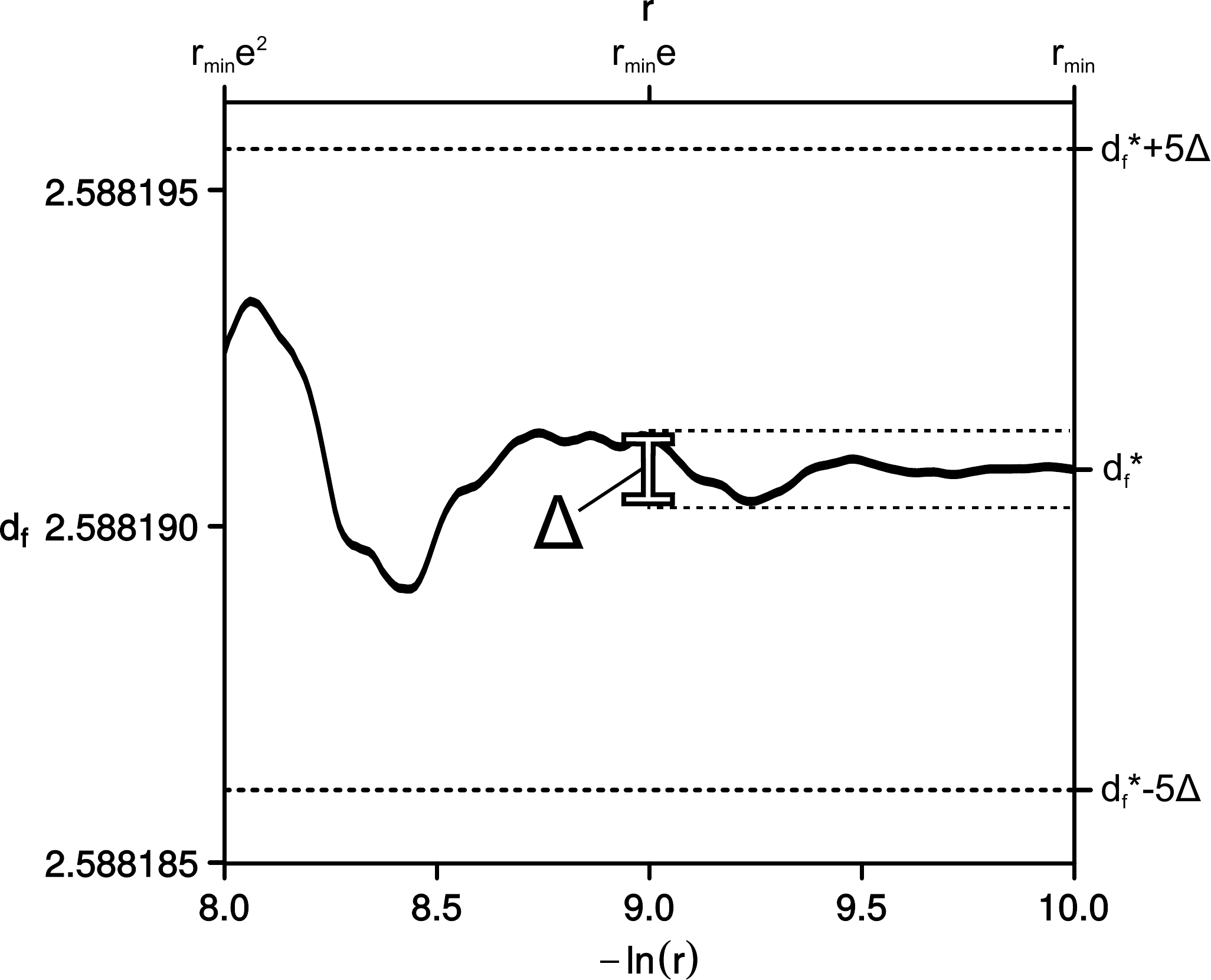}
\end{center}
\caption{
\label{fig:df_best_estimate} As our best estimate (here $d_f^*$), we take of all estimate functions (compare Fig.\ \ref{fig:df_estimates}), the one that shows the least variability $\Delta$ in the radius interval $[r_\text{min},r_\text{min}e]$ (here the function based on the combination of $s$ and $p$ from Fig.\ \ref{fig:df_estimates}), and take the confidence interval $\hat{d}_f (r_\text{min}) \pm 5 \Delta$ as our best estimate. This is the previously known 3D packing of Ref.\ \cite{Baram2004}.
}
\end{figure}

\begin{figure}[]
\begin{center}
	\includegraphics[width=\columnwidth]{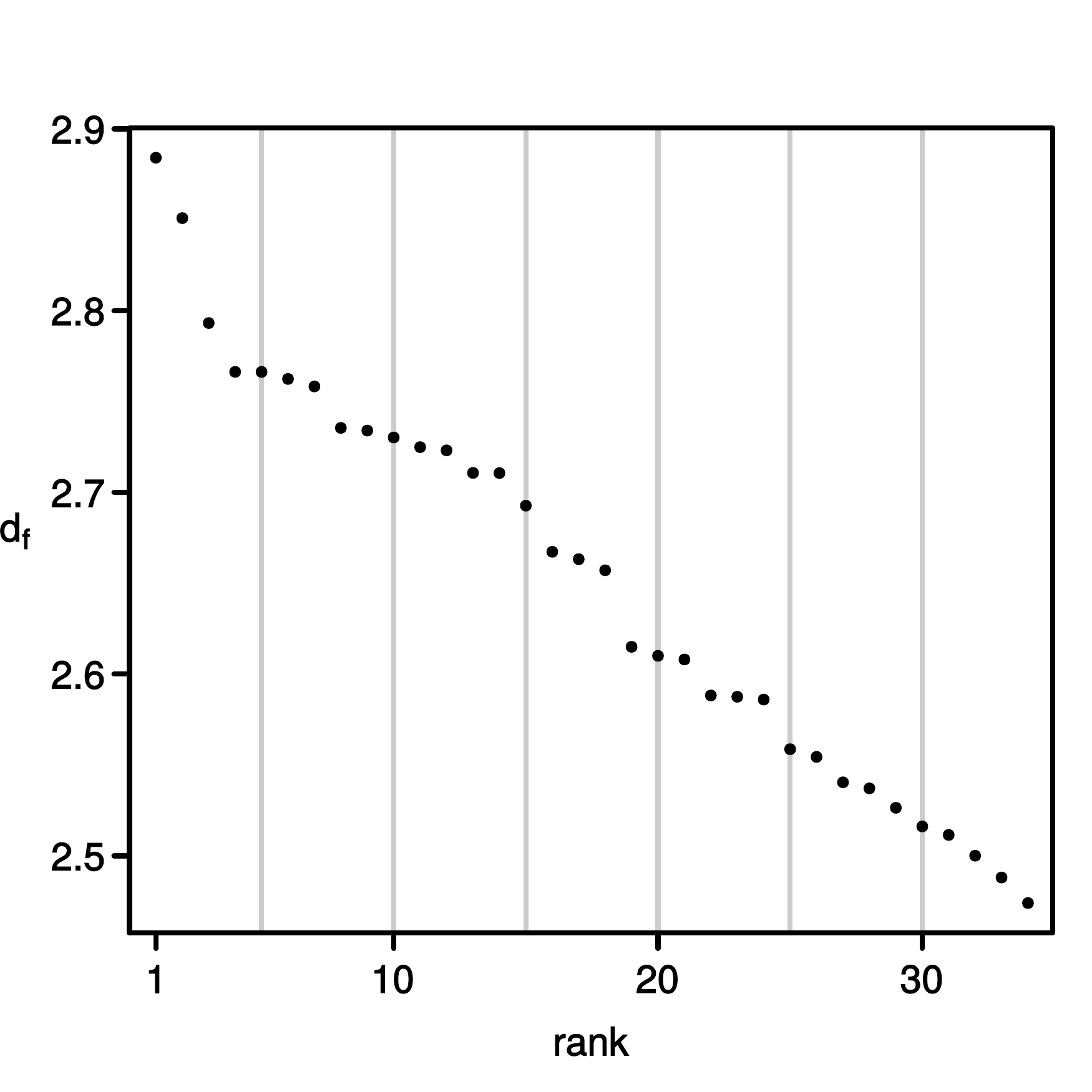}
\end{center}
\caption{
\label{fig:dfs_3D} Ranked fractal dimension estimates of 3D packings. The calculated confidence intervals are smaller than the symbol sizes.
}
\end{figure}

\begin{figure}[]
\begin{center}
	\includegraphics[width=\columnwidth]{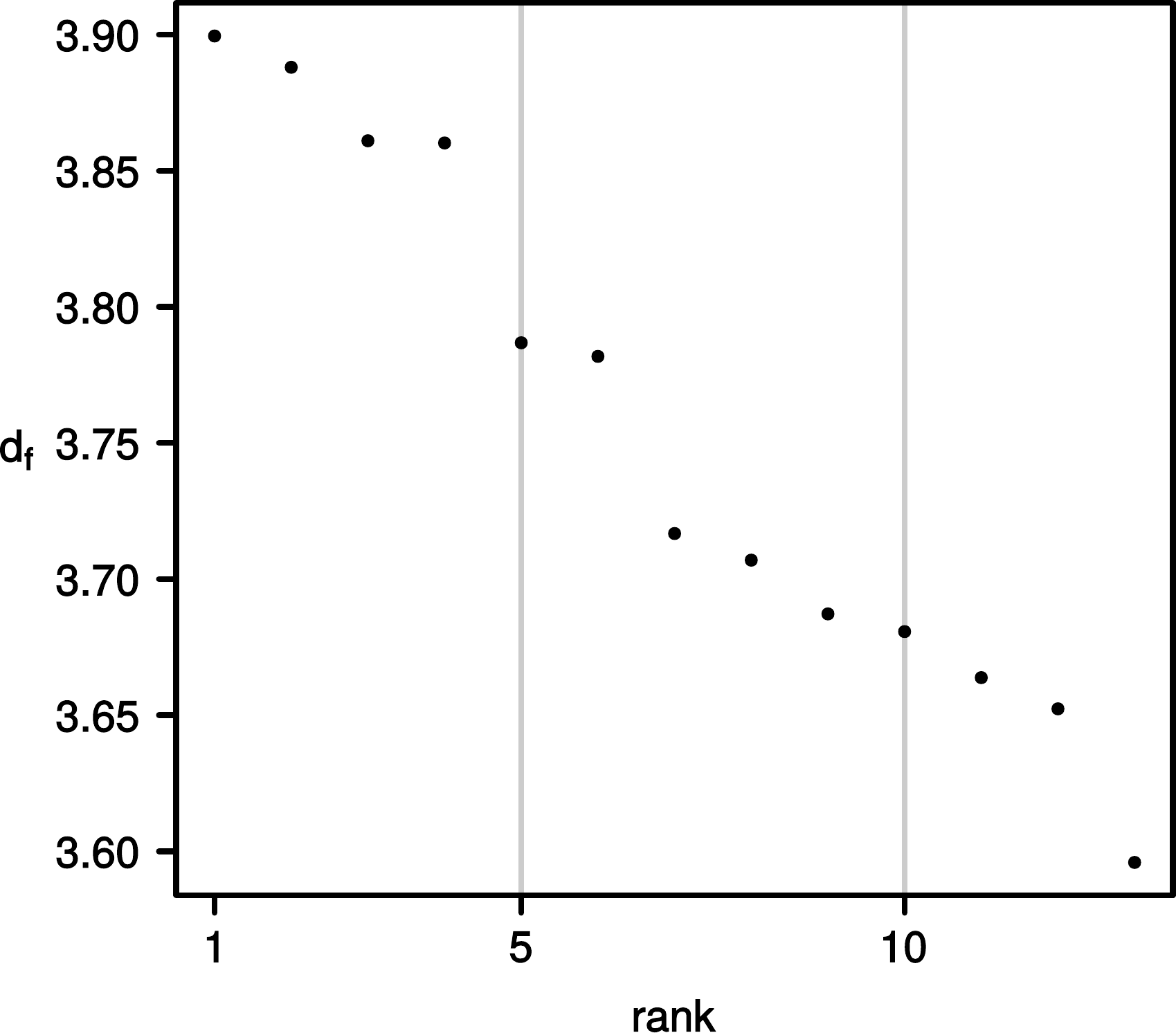}
\end{center}
\caption{
\label{fig:dfs_4D} Ranked fractal dimension estimates of 4D packings. The calculated confidence intervals are smaller than the symbol sizes.
}
\end{figure}

Finally, we take from all estimate functions $\hat{d}_f (r)$ the one with the least variability $\Delta$ in the interval $[r_\text{min},r_\text{min}e]$, and take $\hat{d}_f (r_\text{min}) \pm 5 \Delta$ as our confidence interval for our best estimation as shown in Fig.\ \ref{fig:df_best_estimate}. The lowest considered radius was $r_\text{min}=e^{-10}$ and $r_\text{min}=e^{-7}$ for 3D and 4D, respectively. Our best estimates for all our 3D and 4D packings can be found in decreasing order in Figs.\ \ref{fig:dfs_3D} and \ref{fig:dfs_4D}, and in the summarizing Tables \ref{tab:3D} and \ref{tab:4D}. For one of the previously known packings, the 3D Apollonian Gasket, we show the more accurate estimation from Ref.\ \cite{Borkovec1994}, which lays within our determined confidence interval.

\subsection{Contact network}\label{sec:clusters}

Depending on the spatial arrangement of the seeds and inversion spheres in the generating setup, the resulting packing forms a single or multiple clusters, where a cluster is a connected network of touching spheres. In particular, non contacting seeds can lead to isolated spheres that are not in contact with any other spheres.

\begin{figure}[]
\begin{center}
	\includegraphics[width=\columnwidth]{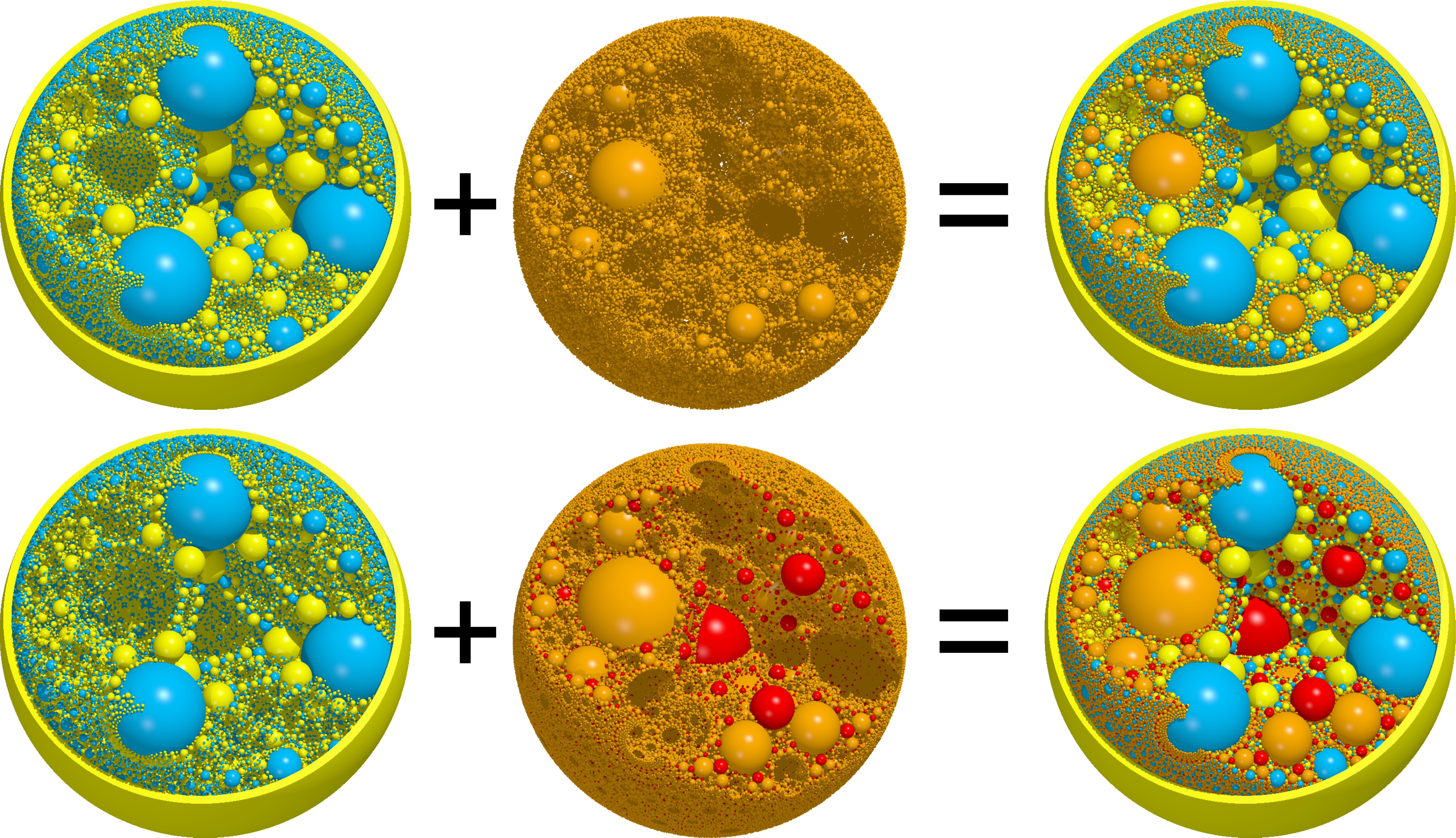}
\end{center}
\caption{
\label{fig:clusters} Two packings with different contact networks. Top: packing consisting of a bipartite cluster with smallest loop size six (left) and isolated spheres (middle). Bottom: packing consisting of two clusters, a bipartite one with smallest loop size twelve (left) and a non-bipartite one with smallest loop size three (middle).
}
\end{figure}

We characterized the contact network of each packing by counting the number of connected clusters, determine the size of their smallest loops, check if clusters are bipartite, and finally check for isolated spheres, as explained in detail in the following paragraphs. Figure \ref{fig:clusters} shows two packings with different contact networks. All results can be found in the summarizing Tables \ref{tab:3D} and \ref{tab:4D}. We found that most topologies only have one connected cluster (31 of 34 in 3D and 10 of 13 in 4D). Besides, there are 3 two-cluster topologies in 3D, and 2 two-cluster topologies and even one with three connected clusters in 4D. Isolated spheres are present only in topologies with a single bipartite cluster (in 7 of 34 in 3D and in 3 of 13 in 4D). Most clusters are bipartite (62\% in 3D and 88\% in 4D). Note that in addition to a single previously known exactly self-similar bearing, which here we specify as a topology merely consisting of a single bipartite cluster, we found another 10 and 5 bearings in 3D and 4D, respectively.

The number of connected clusters can be derived directly from the generating setup. Each cluster of touching seeds will lead to a connected cluster in the packing. An isolated seed will lead to isolated spheres in the packing only if it is not tangent to any inversion sphere. Otherwise, it will form a cluster with its inversions.

To find the smallest loop size of clusters, we generated each packing up to a certain generation of spheres, i.e., up to a certain number of successive inversions starting from the seeds, such that every seed is part of a closed loop. Every loop can be mapped onto a loop where all pairwise touching spheres are at most one generation apart. Therefore, by iterative inversion of all seeds and their successive images, the smallest loop that a certain seed is part of will be closed first during the generation process. For every seed we find the first closed and therefore smallest loop that it belongs to, and from that we derive the smallest loops of the clusters.

If all smallest loops of different seeds of a cluster are even, the cluster is bipartite, since one can divide the generated spheres into two groups A and B, such that A-spheres only touch B-spheres and vice versa. Then, every contact that appears if more spheres are generated can be mapped onto a contact between a previously generated A-sphere and B-sphere.

\begin{table*}[]
\centering
\caption{Summary of 3D generating setups. Base shape of each setup is a Platonic Solid with number of faces shown in column B. The position of the outer inversion spheres is given in column I and can be either the edges (E) or the faces (F) of the base shape. Family number F and parameter b and c in corresponding columns. Smallest loop size of clusters with letter (b) for bipartite clusters in column ``loop size''. Presence of isolated spheres indicated by cross-marks in column ``isolated''. Estimate of fractal dimension and rank R in comparison with all packings (R=1 for highest fractal dimension) in corresponding columns. Some setups lead to identical topologies which are assigned a reference number in column T, where a (*) signifies that we assume the topology to be equal to another not because the generated packings are identical but we can not topologically distinguish them. Bearings, i.e., packings merely consisting of a single bipartite cluster, have a cross-mark in column ``bearing''. Last column gives information about previous knowledge of the packings.}
\label{tab:3D}
\begin{tabular}{|l|l|l|l|l|l|l|l|l|}
\hline
\textbf{b} & \textbf{c} & \textbf{loop size} & \textbf{isolated} & \textbf{fractal dimension} & \textbf{R} & \textbf{T} & \textbf{bearing} & \textbf{previously known as}                                         \\ \hline
0          & 0          & 3                  &                   & 2.52638$\pm$0.000005       & 29         & 1          &                  & Cube based  packing of Ref.\ \cite{Baram2004a}        \\ \hline
0          & 1          & 3                  &                   & 2.53706$\pm$0.00002        & 28         &            &                  &                                                                      \\ \hline
0          & 2          & 3                  &                   & 2.51613$\pm$0.00002        & 30         &            &                  &                                                                      \\ \hline
0          & 3          & 3                  &                   & 2.50002$\pm$0.00006        & 32         &            &                  &                                                                      \\ \hline
0          & 4          & 3                  &                   & 2.4739465$\pm$0.0000001    & 34         & 2          &                  & Apollonian Gasket                                                    \\ \hline
0          & 0          & 3                  &                   & 2.55863$\pm$0.00002        & 25         & 3          &                  &                                                                      \\ \hline
0          & 1          & 3                  &                   & 2.55438$\pm$0.00008        & 26         &            &                  &                                                                      \\ \hline
0          & 0          & 3                  &                   & 2.4739465$\pm$0.0000001    & 34         & 2          &                  & Apollonian Gasket                                                    \\ \hline
0          & 1          & 4b                 &                   & 2.71066$\pm$0.00002        & 13         &            & \checkmark       &                                                                      \\ \hline
0          & 2          & 4b                 &                   & 2.76236$\pm$0.00005        & 6          & 4*         & \checkmark       &                                                                      \\ \hline
0          & 3          & 4b                 & \checkmark        & 2.76625$\pm$0.00005        & 4          &            &                  &                                                                      \\ \hline
0          & 4          & 4b                 &                   & 2.692627$\pm$0.000002      & 15         &            & \checkmark       &                                                                      \\ \hline
0          & 0          & 3                  &                   & 2.4739465$\pm$0.0000001    & 34         & 2          &                  & Apollonian Gasket                                                    \\ \hline
0          & 1          & 4b                 &                   & 2.73543$\pm$0.0001         & 8          &            & \checkmark       &                                                                      \\ \hline
0          & 2          & 4b                 & \checkmark        & 2.72307$\pm$0.00009        & 12         & 5*         &                  &                                                                      \\ \hline
0          & 3          & 4b                 & \checkmark        & 2.66723$\pm$0.00005        & 16         & 6*         &                  &                                                                      \\ \hline
0          & 4          & 3                  &                   & 2.61$\pm$0.0002            & 20         &            &                  &                                                                      \\ \hline
0          & 0          & 3                  &                   & 2.52638$\pm$0.000005       & 29         & 1          &                  & Cube based packing of Ref.\ \cite{Baram2004a}         \\ \hline
0          & 1          & 4b                 &                   & 2.76235$\pm$0.00003        & 6          & 4*         & \checkmark       &                                                                      \\ \hline
0          & 2          & 4b,12b             &                   & 2.75823$\pm$0.00003        & 7          & 7*         &                  &                                                                      \\ \hline
0          & 0          & 3                  &                   & 2.55863$\pm$0.00002        & 25         & 3          &                  &                                                                      \\ \hline
0          & 1          & 4b                 & \checkmark        & 2.723057$\pm$0.000003      & 12         & 5*         &                  &                                                                      \\ \hline
0          & 2          & 4b,6b              &                   & 2.65707$\pm$0.00002        & 18         & 8*         &                  &                                                                      \\ \hline
0          & 0          & 4b                 &                   & 2.61496$\pm$0.00002        & 19         & 9*         & \checkmark       &                                                                      \\ \hline
0          & 1          & 6b                 & \checkmark        & 2.73397$\pm$0.00002        & 9          & 10*        &                  &                                                                      \\ \hline
0          & 2          & 8b                 & \checkmark        & 2.71055$\pm$0.00004        & 14         & 11*        &                  &                                                                      \\ \hline
0          & 3          & 10b                & \checkmark        & 2.66319$\pm$0.00003        & 17         & 12*        &                  &                                                                      \\ \hline
0          & 4          & 3,12b              &                   & 2.60799$\pm$0.00006        & 21         & 13*        &                  &                                                                      \\ \hline
1          & 0          & 4b                 &                   & 2.588191$\pm$0.000005      & 22         & 14         & \checkmark       & Ref.\ \cite{Baram2004}                                \\ \hline
1          & 1          & 4b                 &                   & 2.61496$\pm$0.00002        & 19         & 9*         & \checkmark       &                                                                      \\ \hline
1          & 2          & 3                  &                   & 2.58747$\pm$0.00002        & 23         &            &                  &                                                                      \\ \hline
0          & 0          & 4b                 &                   & 2.730156$\pm$0.000005      & 10         & 15         & \checkmark       &                                                                      \\ \hline
0          & 1          & 4b,6b              &                   & 2.65707$\pm$0.00002        & 18         & 8*         &                  &                                                                      \\ \hline
1          & 0          & 4b                 &                   & 2.588191$\pm$0.000005      & 22         & 14         & \checkmark       & Ref.\ \cite{Baram2004}                                \\ \hline
0          & 0          & 4b                 &                   & 2.588191$\pm$0.000005      & 22         & 14         & \checkmark       & Ref.\ \cite{Baram2004}                                \\ \hline
0          & 1          & 6b                 &                   & 2.793143$\pm$0.000005      & 3          &            & \checkmark       &                                                                      \\ \hline
0          & 2          & 8b                 &                   & 2.8841$\pm$0.00005         & 1          &            & \checkmark       &                                                                      \\ \hline
0          & 3          & 10b                & \checkmark        & 2.850875$\pm$0.000007      & 2          &            &                  &                                                                      \\ \hline
0          & 4          & 4b,12b             &                   & 2.75824$\pm$0.00004        & 7          & 7*         &                  &                                                                      \\ \hline
1          & 0          & 3                  &                   & 2.488006$\pm$0.000008      & 33         & 16         &                  & octahedron based packing of Ref.\ \cite{Baram2004a}   \\ \hline
1          & 1          & 4b                 &                   & 2.724834$\pm$0.000009      & 11         &            & \checkmark       &                                                                      \\ \hline
1          & 2          & 4b                 &                   & 2.730156$\pm$0.000005      & 10         & 15         & \checkmark       &                                                                      \\ \hline
0          & 0          & 4b                 &                   & 2.61496$\pm$0.00002        & 19         & 9*         & \checkmark       &                                                                      \\ \hline
0          & 1          & 6b                 & \checkmark        & 2.73398$\pm$0.00005        & 9          & 10*        &                  &                                                                      \\ \hline
0          & 2          & 8b                 & \checkmark        & 2.71055$\pm$0.00008        & 14         & 11*        &                  &                                                                      \\ \hline
0          & 3          & 10b                & \checkmark        & 2.66319$\pm$0.00006        & 17         & 12*        &                  &                                                                      \\ \hline
0          & 4          & 3,12b              &                   & 2.608$\pm$0.0002           & 21         & 13*        &                  &                                                                      \\ \hline
1          & 0          & 3                  &                   & 2.488006$\pm$0.000008      & 33         & 16         &                  & octahedron based packing of Ref.\ \cite{Baram2004a}   \\ \hline
1          & 1          & 4b                 &                   & 2.588191$\pm$0.000005      & 22         & 14         & \checkmark       & Ref.\ \cite{Baram2004}                                \\ \hline
1          & 2          & 3                  &                   & 2.5404$\pm$0.0003          & 27         &            &                  &                                                                      \\ \hline
0          & 0          & 3                  &                   & 2.51142$\pm$0.00004        & 31         &            &                  & dodecahedron based packing of Ref.\ \cite{Baram2004a} \\ \hline
0          & 1          & 4b                 &                   & 2.76624$\pm$0.00004        & 5          &            & \checkmark       &                                                                      \\ \hline
0          & 0          & 3                  &                   & 2.58594$\pm$0.00002        & 24         &            &                  &                                                                      \\ \hline
0          & 1          & 4b                 & \checkmark        & 2.66722$\pm$0.00002        & 16         & 6*         &                  &                                                                      \\ \hline
\end{tabular}
\end{table*}

\begin{table*}[]
\centering
\caption{Summary of 4D generating setups. Base shape of each setup is a convex regular 4-polytope with number of cells shown in column B. The position of the outer inversion spheres is given in column I and can be either the edges (E), faces (F), or cells (C) of the base shape. Family number F and parameter b and c in corresponding columns. Smallest loop size of clusters with letter (b) for bipartite clusters in column ``loop size''. Presence of isolated spheres in column ``isolated''. Estimate of fractal dimension $d_f$ and rank R in comparison with all packings (R=1 for highest fractal dimension) in corresponding columns. Some setups lead to identical topologies which are assigned a reference number in column T, where a (*) signifies that we assume the topology to be equal to another not because the generated packings are identical but we can not topologically distinguish them. Bearings, i.e., packings merely consisting of a single bipartite cluster, indicated in column ``bearing''.}
\label{tab:4D}
\begin{tabular}{|l|l|l|l|l|l|l|l|l|l|l|}
\hline
\textbf{B} & \textbf{I} & \textbf{F} & \textbf{b} & \textbf{c} & \textbf{loop size} & \textbf{isolated} & \textbf{fractal dimension} & \textbf{R} & \textbf{T} & \textbf{bearing} \\ \hline
5          & F          & 1          & 0          & 0          & 4b                 &                   & 3.6807$\pm$0.0009          & 10         & 1*         & \checkmark       \\ \hline
5          & F          & 1          & 0          & 1          & 6b                 & \checkmark        & 3.7818$\pm$0.0009          & 6          & 2*         &                  \\ \hline
5          & F          & 1          & 0          & 2          & 4b,8b              &                   & 3.6872$\pm$0.0003          & 9          & 3*         &                  \\ \hline
5          & F          & 2          & 0          & 0          & 3                  &                   & 3.59591$\pm$0.00002        & 13         & 4*         &                  \\ \hline
5          & F          & 2          & 0          & 1          & 4b                 &                   & 3.70695$\pm$0.00002        & 8          & 5*         & \checkmark       \\ \hline
8          & F          & 1          & 0          & 0          & 4b                 &                   & 3.66379$\pm$0.00004        & 11         & 6          & \checkmark       \\ \hline
8          & F          & 1          & 0          & 1          & 4b                 &                   & 3.6807$\pm$0.001           & 10         & 1*         & \checkmark       \\ \hline
8          & F          & 2          & 0          & 0          & 3                  &                   & 3.65233$\pm$0.00005        & 12         & 7          &                  \\ \hline
16         & E          & 1          & 0          & 0          & 4b                 &                   & 3.66379$\pm$0.00004        & 11         & 6          & \checkmark       \\ \hline
16         & E          & 2          & 0          & 0          & 3                  &                   & 3.65233$\pm$0.00005        & 12         & 7          &                  \\ \hline
16         & F          & 1          & 0          & 0          & 4b                 &                   & 3.70695$\pm$0.00003        & 8          & 5          & \checkmark       \\ \hline
16         & F          & 1          & 0          & 1          & 6b                 &                   & 3.8995$\pm$0.001           & 1          &            & \checkmark       \\ \hline
16         & F          & 1          & 0          & 2          & 8b                 & \checkmark        & 3.8602$\pm$0.0004          & 4          &            &                  \\ \hline
16         & F          & 1          & 1          & 0          & 3                  &                   & 3.59591$\pm$0.00003        & 13         & 4          &                  \\ \hline
16         & F          & 1          & 1          & 1          & 4b                 & \checkmark        & 3.7868$\pm$0.0006          & 5          & 8*         &                  \\ \hline
16         & F          & 2          & 0          & 0          & 3                  &                   & 3.59591$\pm$0.00003        & 13         & 4          &                  \\ \hline
16         & F          & 2          & 0          & 1          & 4b                 &                   & 3.70695$\pm$0.00003        & 8          & 5          & \checkmark       \\ \hline
24         & F          & 1          & 0          & 0          & 4b                 &                   & 3.6806$\pm$0.0002          & 10         & 1*         & \checkmark       \\ \hline
24         & F          & 1          & 0          & 1          & 6b                 & \checkmark        & 3.7816$\pm$0.0003          & 6          & 2*         &                  \\ \hline
24         & F          & 1          & 0          & 2          & 4b,8b              &                   & 3.6872$\pm$0.0004          & 9          & 3*         &                  \\ \hline
24         & F          & 1          & 1          & 0          & 4b                 &                   & 3.66379$\pm$0.00004        & 11         & 6          & \checkmark       \\ \hline
24         & F          & 1          & 1          & 1          & 4b                 &                   & 3.6807$\pm$0.0003          & 10         & 1*         & \checkmark       \\ \hline
24         & F          & 2          & 0          & 0          & 4b                 &                   & 3.66379$\pm$0.00004        & 11         & 6          & \checkmark       \\ \hline
24         & C          & 1          & 0          & 0          & 4b                 &                   & 3.66379$\pm$0.00004        & 11         & 6          & \checkmark       \\ \hline
24         & C          & 1          & 0          & 1          & 6b                 &                   & 3.888$\pm$0.002            & 2          &            & \checkmark       \\ \hline
24         & C          & 1          & 0          & 2          & 8b,8b,8b           &                   & 3.861$\pm$0.0006           & 3          &            &                  \\ \hline
24         & C          & 2          & 0          & 0          & 3                  &                   & 3.65233$\pm$0.00005        & 12         & 7          &                  \\ \hline
24         & C          & 2          & 0          & 1          & 4b                 & \checkmark        & 3.787$\pm$0.002            & 5          & 8*         &                  \\ \hline
24         & C          & 2          & 0          & 2          & 4b,4b              &                   & 3.71673$\pm$0.00005        & 7          &            &                  \\ \hline
\end{tabular}
\end{table*}

\section{Modified packings}\label{sec:modify_packings}

\begin{figure}[]
\begin{center}
	\includegraphics[width=\columnwidth]{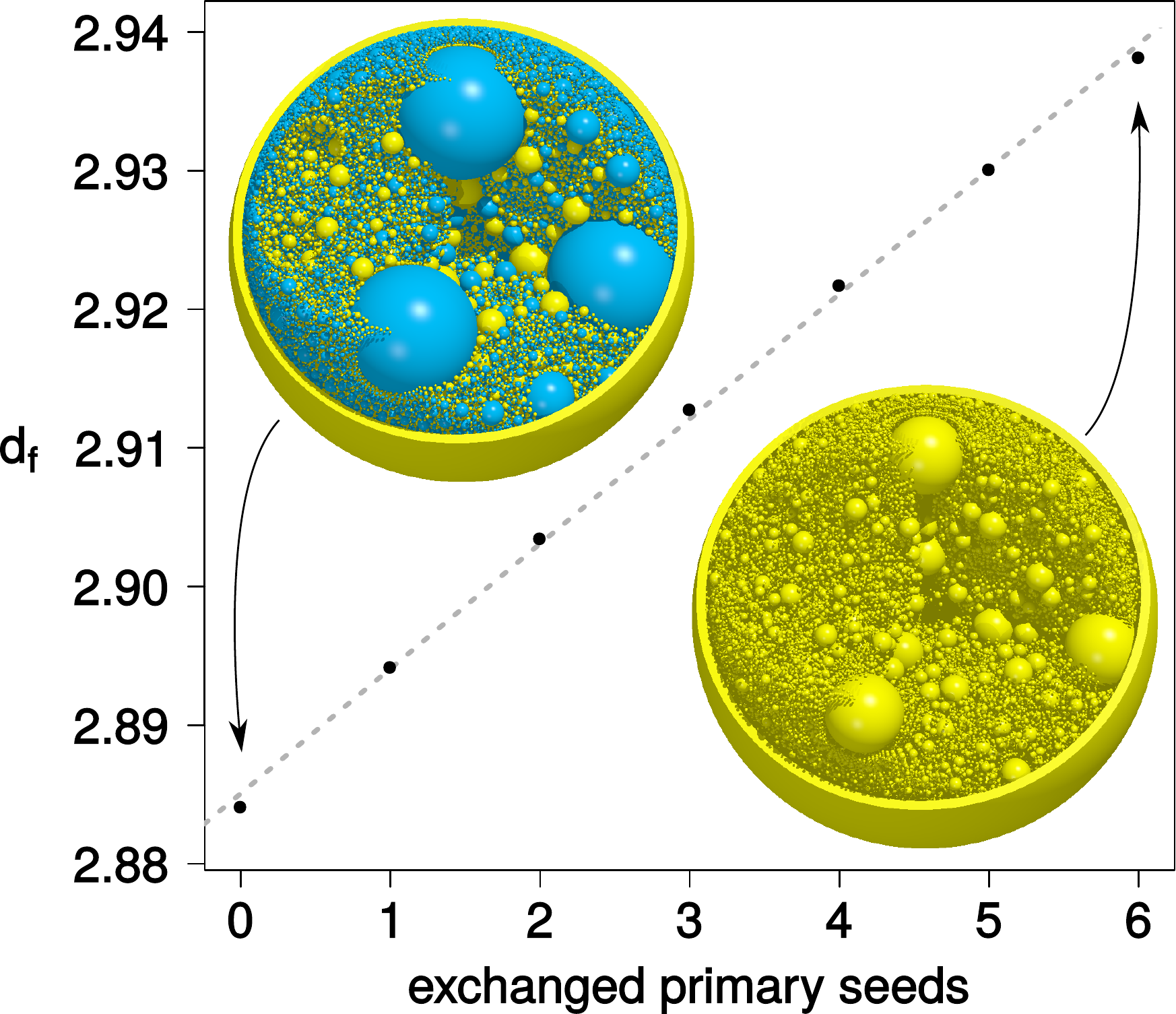}
\end{center}
\caption{
\label{fig:modify_packing} Exchanging seeds by inversion spheres leads to an increased fractal dimension of the resulting packing. At last, the unit sphere hole is the only remaining seed. Confidence intervals are smaller than the size of the symbols. Dashed line is a linear fit.
}
\end{figure}

Packings can be modified in different ways. First of all, any packing can be inverted as a whole at any inversion sphere, which does not change its topology and therefore neither its fractal dimension, but its symmetry and spatial arrangement. Second, any packing can be nested in another packing by exchaning any sphere of a given packing with a packing that is enclosed by a sphere. Third, one can cut any $n$-dimensional packing with an $m$-dimensional subspace, given $2\leq m < n$, to obtain an $m$-dimensional packing. And fourth, one can exchange seeds of a generating setup with inversion spheres to increase the fractal dimension of the generated packing. As long as a single seed remains, the resulting packing will be self-similar and space-filling. Figure \ref{fig:modify_packing} shows the fractal dimensions of different packings resulting from a setup with different number of primary seeds exchanged with inversion spheres.

\section{Final Remarks}\label{sec:conclusion}

We presented an approach to find setups to generate self-similar space-filling sphere packings in arbitrary dimensions. This allows to generate 34, of which 5 were previously known, and 13 new topologies in 3D and 4D, respectively. We characterized all topologies according to their fractal dimensions and the properties of their contact network. The fractal dimensions range from $2.47$ to $2.88$ in 3D and from $3.60$ to $3.90$ in 4D. We explained how the fractal dimension of the generated packing can be increased by exchanging seeds of the generating setups with inversion spheres. Furthermore, each packing can be cut in various ways to serve as a source of lower dimensional cuts. The presented topologies together with the possible modifications offer the possibility to obtain space-filling packings with various fractal dimensions and contact-network properties.

Reference \cite{Baram2004a} suggests, that self-similar space-filling packings are inhomogenous fractals, such that different cuts of a certain dimension can have different fractal dimensions. A detailed investigation on the fractal dimension of such cuts will be subject of future studies.

We have seen that exchanging seeds with inversion spheres increases the fractal dimension of the resulting packing. Apart from that, it remains an open question how the specifics of the generating setup influences the fractal dimension. A systematic study might allow a more precise search for a generating setup that lead to a certain desired fractal dimension.

Previously, the only known 3D bearings, i.e., single cluster bipartite space-filling packings without isolated spheres, were the exactly self-similar one from Ref.\ \cite{Baram2004} and the random ones from Ref.\ \cite{Baram2005}, which both have smallest loop size four. We found additional ones of which two have a smallest loop size of larger than four, namely the packings based on the octahedron with outer inversion spheres at the faces belonging to F1 with $b=0$ with smallest loop size six ($c=1$) and eight ($c=2$). As shown in Ref.\ \cite{Stager2016}, bipartite space-filling packings with smallest loop size four have slip-free rotation states with four degrees of freedom. When inverted to end up in a packing bounded by two infinite spheres, i.e., bounded by two parallel planes, they allow the simultaneous and synchronized motion of the two parallel planes in any direction. According to Ref.\ \cite{Stager2016}, a smallest loop size of larger than four might lead to more degrees of freedom and extended frictionless functionality, what could be investigated in future studies.

\begin{acknowledgments}
We acknowledge financial support from the ETH Risk Center, the Brazilian institute INCT-SC, Grant No.\ FP7-319968-FlowCCS of the European Research Council (ERC) Advanced Grant. Special thanks goes to Sergio Solorzano Rocha for his encouragement to explore the fourth dimension.
\end{acknowledgments}

\bibliography{bibliography_paper}

\begin{thebibliography}{35}%
\makeatletter
\providecommand \@ifxundefined [1]{%
 \@ifx{#1\undefined}
}%
\providecommand \@ifnum [1]{%
 \ifnum #1\expandafter \@firstoftwo
 \else \expandafter \@secondoftwo
 \fi
}%
\providecommand \@ifx [1]{%
 \ifx #1\expandafter \@firstoftwo
 \else \expandafter \@secondoftwo
 \fi
}%
\providecommand \natexlab [1]{#1}%
\providecommand \enquote  [1]{``#1''}%
\providecommand \bibnamefont  [1]{#1}%
\providecommand \bibfnamefont [1]{#1}%
\providecommand \citenamefont [1]{#1}%
\providecommand \href@noop [0]{\@secondoftwo}%
\providecommand \href [0]{\begingroup \@sanitize@url \@href}%
\providecommand \@href[1]{\@@startlink{#1}\@@href}%
\providecommand \@@href[1]{\endgroup#1\@@endlink}%
\providecommand \@sanitize@url [0]{\catcode `\\12\catcode `\$12\catcode
  `\&12\catcode `\#12\catcode `\^12\catcode `\_12\catcode `\%12\relax}%
\providecommand \@@startlink[1]{}%
\providecommand \@@endlink[0]{}%
\providecommand \url  [0]{\begingroup\@sanitize@url \@url }%
\providecommand \@url [1]{\endgroup\@href {#1}{\urlprefix }}%
\providecommand \urlprefix  [0]{URL }%
\providecommand \Eprint [0]{\href }%
\providecommand \doibase [0]{http://dx.doi.org/}%
\providecommand \selectlanguage [0]{\@gobble}%
\providecommand \bibinfo  [0]{\@secondoftwo}%
\providecommand \bibfield  [0]{\@secondoftwo}%
\providecommand \translation [1]{[#1]}%
\providecommand \BibitemOpen [0]{}%
\providecommand \bibitemStop [0]{}%
\providecommand \bibitemNoStop [0]{.\EOS\space}%
\providecommand \EOS [0]{\spacefactor3000\relax}%
\providecommand \BibitemShut  [1]{\csname bibitem#1\endcsname}%
\let\auto@bib@innerbib\@empty
\bibitem [{\citenamefont {Ayer}\ and\ \citenamefont {Soppet}(1965)}]{Ayer1965}%
  \BibitemOpen
  \bibfield  {author} {\bibinfo {author} {\bibfnamefont {J.~E.}\ \bibnamefont
  {Ayer}}\ and\ \bibinfo {author} {\bibfnamefont {F.~E.}\ \bibnamefont
  {Soppet}},\ }\href@noop {} {\bibfield  {journal} {\bibinfo  {journal} {J. Am.
  Ceram. Soc.}\ }\textbf {\bibinfo {volume} {48}},\ \bibinfo {pages} {180}
  (\bibinfo {year} {1965})}\BibitemShut {NoStop}%
\bibitem [{\citenamefont {Jodrey}\ and\ \citenamefont
  {Tory}(1985)}]{Jodrey1985}%
  \BibitemOpen
  \bibfield  {author} {\bibinfo {author} {\bibfnamefont {W.~S.}\ \bibnamefont
  {Jodrey}}\ and\ \bibinfo {author} {\bibfnamefont {E.~M.}\ \bibnamefont
  {Tory}},\ }\href {\doibase 10.1103/PhysRevA.32.2347} {\bibfield  {journal}
  {\bibinfo  {journal} {Phys. Rev. A}\ }\textbf {\bibinfo {volume} {32}},\
  \bibinfo {pages} {2347} (\bibinfo {year} {1985})}\BibitemShut {NoStop}%
\bibitem [{\citenamefont {Yu}\ and\ \citenamefont {Standish}(1988)}]{Yu1988}%
  \BibitemOpen
  \bibfield  {author} {\bibinfo {author} {\bibfnamefont {A.~B.}\ \bibnamefont
  {Yu}}\ and\ \bibinfo {author} {\bibfnamefont {N.}~\bibnamefont {Standish}},\
  }\href {\doibase 10.1016/0032-5910(88)80101-3} {\bibfield  {journal}
  {\bibinfo  {journal} {Powder Technol.}\ }\textbf {\bibinfo {volume} {55}},\
  \bibinfo {pages} {171} (\bibinfo {year} {1988})}\BibitemShut {NoStop}%
\bibitem [{\citenamefont {Ouchiyama}\ and\ \citenamefont
  {Tanaka}(1989)}]{Ouchiyama1989}%
  \BibitemOpen
  \bibfield  {author} {\bibinfo {author} {\bibfnamefont {N.}~\bibnamefont
  {Ouchiyama}}\ and\ \bibinfo {author} {\bibfnamefont {T.}~\bibnamefont
  {Tanaka}},\ }\href {\doibase 10.1021/ie00094a016} {\bibfield  {journal}
  {\bibinfo  {journal} {Ind. Eng. Chem. Res.}\ }\textbf {\bibinfo {volume}
  {28}},\ \bibinfo {pages} {1530} (\bibinfo {year} {1989})}\BibitemShut
  {NoStop}%
\bibitem [{\citenamefont {Soppe}(1990)}]{Soppe1990}%
  \BibitemOpen
  \bibfield  {author} {\bibinfo {author} {\bibfnamefont {W.}~\bibnamefont
  {Soppe}},\ }\href@noop {} {\ \textbf {\bibinfo {volume} {62}},\ \bibinfo
  {pages} {189} (\bibinfo {year} {1990})}\BibitemShut {NoStop}%
\bibitem [{\citenamefont {Konakawa}\ and\ \citenamefont
  {Ishizaki}(1990)}]{Konakawa1990}%
  \BibitemOpen
  \bibfield  {author} {\bibinfo {author} {\bibfnamefont {Y.}~\bibnamefont
  {Konakawa}}\ and\ \bibinfo {author} {\bibfnamefont {K.}~\bibnamefont
  {Ishizaki}},\ }\href {\doibase 10.1016/0032-5910(90)80049-5} {\bibfield
  {journal} {\bibinfo  {journal} {Powder Technol.}\ }\textbf {\bibinfo {volume}
  {63}},\ \bibinfo {pages} {241} (\bibinfo {year} {1990})}\BibitemShut
  {NoStop}%
\bibitem [{\citenamefont {Standish}\ \emph {et~al.}(1991)\citenamefont
  {Standish}, \citenamefont {Yu},\ and\ \citenamefont {Zou}}]{Standish1991}%
  \BibitemOpen
  \bibfield  {author} {\bibinfo {author} {\bibfnamefont {N.}~\bibnamefont
  {Standish}}, \bibinfo {author} {\bibfnamefont {A.~B.}\ \bibnamefont {Yu}}, \
  and\ \bibinfo {author} {\bibfnamefont {R.~P.}\ \bibnamefont {Zou}},\ }\href
  {\doibase 10.1016/0032-5910(91)80126-4} {\bibfield  {journal} {\bibinfo
  {journal} {Powder Technol.}\ }\textbf {\bibinfo {volume} {68}},\ \bibinfo
  {pages} {175} (\bibinfo {year} {1991})}\BibitemShut {NoStop}%
\bibitem [{\citenamefont {Yu}\ and\ \citenamefont {Standish}(1993)}]{Yu1993}%
  \BibitemOpen
  \bibfield  {author} {\bibinfo {author} {\bibfnamefont {A.~B.}\ \bibnamefont
  {Yu}}\ and\ \bibinfo {author} {\bibfnamefont {N.}~\bibnamefont {Standish}},\
  }\href {\doibase 10.1016/S0032-5910(05)80018-X} {\bibfield  {journal}
  {\bibinfo  {journal} {Powder Technol.}\ }\textbf {\bibinfo {volume} {76}},\
  \bibinfo {pages} {113} (\bibinfo {year} {1993})}\BibitemShut {NoStop}%
\bibitem [{\citenamefont {Anishchik}\ and\ \citenamefont
  {Medvedev}(1995)}]{Anishchik1995}%
  \BibitemOpen
  \bibfield  {author} {\bibinfo {author} {\bibfnamefont {S.~V.}\ \bibnamefont
  {Anishchik}}\ and\ \bibinfo {author} {\bibfnamefont {N.~N.}\ \bibnamefont
  {Medvedev}},\ }\href {\doibase 10.1103/PhysRevLett.75.4314} {\bibfield
  {journal} {\bibinfo  {journal} {Phys. Rev. Lett.}\ }\textbf {\bibinfo
  {volume} {75}},\ \bibinfo {pages} {4314} (\bibinfo {year}
  {1995})}\BibitemShut {NoStop}%
\bibitem [{\citenamefont {Elliott}\ \emph {et~al.}(2002)\citenamefont
  {Elliott}, \citenamefont {Kelly},\ and\ \citenamefont
  {Windle}}]{Elliott2002}%
  \BibitemOpen
  \bibfield  {author} {\bibinfo {author} {\bibfnamefont {J.~A.}\ \bibnamefont
  {Elliott}}, \bibinfo {author} {\bibfnamefont {A.}~\bibnamefont {Kelly}}, \
  and\ \bibinfo {author} {\bibfnamefont {A.~H.}\ \bibnamefont {Windle}},\
  }\href {\doibase 10.1023/A:1016503002392} {\bibfield  {journal} {\bibinfo
  {journal} {J. Mater. Sci. Lett.}\ }\textbf {\bibinfo {volume} {21}},\
  \bibinfo {pages} {1249} (\bibinfo {year} {2002})}\BibitemShut {NoStop}%
\bibitem [{\citenamefont {Sobolev}\ and\ \citenamefont
  {Amirjanov}(2010)}]{Sobolev2010}%
  \BibitemOpen
  \bibfield  {author} {\bibinfo {author} {\bibfnamefont {K.}~\bibnamefont
  {Sobolev}}\ and\ \bibinfo {author} {\bibfnamefont {A.}~\bibnamefont
  {Amirjanov}},\ }\href {\doibase 10.1016/j.conbuildmat.2010.01.010} {\bibfield
   {journal} {\bibinfo  {journal} {Constr. Build. Mater.}\ }\textbf {\bibinfo
  {volume} {24}},\ \bibinfo {pages} {1449} (\bibinfo {year}
  {2010})}\BibitemShut {NoStop}%
\bibitem [{\citenamefont {Rahmani}(2014)}]{Rahmani2014}%
  \BibitemOpen
  \bibfield  {author} {\bibinfo {author} {\bibfnamefont {H.}~\bibnamefont
  {Rahmani}},\ }\href {\doibase 10.1007/s10035-014-0513-5} {\bibfield
  {journal} {\bibinfo  {journal} {Granul. Matter}\ }\textbf {\bibinfo {volume}
  {16}},\ \bibinfo {pages} {751} (\bibinfo {year} {2014})}\BibitemShut
  {NoStop}%
\bibitem [{\citenamefont {Mart{\'{i}}n}\ \emph {et~al.}(2014)\citenamefont
  {Mart{\'{i}}n}, \citenamefont {Mu{\~{n}}os}, \citenamefont {Reyes},\ and\
  \citenamefont {Taguas}}]{Martin2014}%
  \BibitemOpen
  \bibfield  {author} {\bibinfo {author} {\bibfnamefont {M.~A.}\ \bibnamefont
  {Mart{\'{i}}n}}, \bibinfo {author} {\bibfnamefont {F.~J.}\ \bibnamefont
  {Mu{\~{n}}os}}, \bibinfo {author} {\bibfnamefont {M.}~\bibnamefont {Reyes}},
  \ and\ \bibinfo {author} {\bibfnamefont {F.~J.}\ \bibnamefont {Taguas}},\
  }\href {\doibase 10.1142/S0218348X1440009X} {\bibfield  {journal} {\bibinfo
  {journal} {Fractals}\ }\textbf {\bibinfo {volume} {22}},\ \bibinfo {pages}
  {1440009} (\bibinfo {year} {2014})}\BibitemShut {NoStop}%
\bibitem [{\citenamefont {Mart{\'{i}}n}\ \emph {et~al.}(2015)\citenamefont
  {Mart{\'{i}}n}, \citenamefont {Mu{\~{n}}oz}, \citenamefont {Reyes},\ and\
  \citenamefont {Taguas}}]{Martin2015}%
  \BibitemOpen
  \bibfield  {author} {\bibinfo {author} {\bibfnamefont {M.~A.}\ \bibnamefont
  {Mart{\'{i}}n}}, \bibinfo {author} {\bibfnamefont {F.~J.}\ \bibnamefont
  {Mu{\~{n}}oz}}, \bibinfo {author} {\bibfnamefont {M.}~\bibnamefont {Reyes}},
  \ and\ \bibinfo {author} {\bibfnamefont {F.~J.}\ \bibnamefont {Taguas}},\
  }\href {\doibase 10.1007/s00024-014-0918-4} {\bibfield  {journal} {\bibinfo
  {journal} {Pure Appl. Geophys.}\ }\textbf {\bibinfo {volume} {172}},\
  \bibinfo {pages} {141} (\bibinfo {year} {2015})}\BibitemShut {NoStop}%
\bibitem [{\citenamefont {Manna}\ and\ \citenamefont
  {Herrmann}(1991)}]{Manna1991}%
  \BibitemOpen
  \bibfield  {author} {\bibinfo {author} {\bibfnamefont {S.~S.}\ \bibnamefont
  {Manna}}\ and\ \bibinfo {author} {\bibfnamefont {H.~J.}\ \bibnamefont
  {Herrmann}},\ }\href {http://iopscience.iop.org/0305-4470/24/9/006}
  {\bibfield  {journal} {\bibinfo  {journal} {J. Phys. A. Math. Gen.}\ }\textbf
  {\bibinfo {volume} {24}},\ \bibinfo {pages} {L481} (\bibinfo {year}
  {1991})}\BibitemShut {NoStop}%
\bibitem [{\citenamefont {Manna}\ and\ \citenamefont
  {Vicsek}(1991)}]{Manna1991a}%
  \BibitemOpen
  \bibfield  {author} {\bibinfo {author} {\bibfnamefont {S.~S.}\ \bibnamefont
  {Manna}}\ and\ \bibinfo {author} {\bibfnamefont {T.}~\bibnamefont {Vicsek}},\
  }\href {\doibase 10.1007/BF01048305} {\bibfield  {journal} {\bibinfo
  {journal} {J. Stat. Phys.}\ }\textbf {\bibinfo {volume} {64}},\ \bibinfo
  {pages} {525} (\bibinfo {year} {1991})}\BibitemShut {NoStop}%
\bibitem [{\citenamefont {Borkovec}\ \emph {et~al.}(1994)\citenamefont
  {Borkovec}, \citenamefont {Paris},\ and\ \citenamefont
  {Peikert}}]{Borkovec1994}%
  \BibitemOpen
  \bibfield  {author} {\bibinfo {author} {\bibfnamefont {M.}~\bibnamefont
  {Borkovec}}, \bibinfo {author} {\bibfnamefont {W.~D.}\ \bibnamefont {Paris}},
  \ and\ \bibinfo {author} {\bibfnamefont {R.}~\bibnamefont {Peikert}},\ }\href
  {http://www.worldscientific.com/doi/pdf/10.1142/s0218348x94000739} {\bibfield
   {journal} {\bibinfo  {journal} {Fractals}\ }\textbf {\bibinfo {volume}
  {2}},\ \bibinfo {pages} {521} (\bibinfo {year} {1994})}\BibitemShut {NoStop}%
\bibitem [{\citenamefont {Sevier}(1995)}]{Sevier1995}%
  \BibitemOpen
  \bibfield  {author} {\bibinfo {author} {\bibfnamefont {E.~I.}\ \bibnamefont
  {Sevier}},\ }\href@noop {} {\ \textbf {\bibinfo {volume} {86}},\ \bibinfo
  {pages} {113} (\bibinfo {year} {1995})}\BibitemShut {NoStop}%
\bibitem [{\citenamefont {Doye}\ and\ \citenamefont {Massen}(2005)}]{Doye2005}%
  \BibitemOpen
  \bibfield  {author} {\bibinfo {author} {\bibfnamefont {J.~P.~K.}\
  \bibnamefont {Doye}}\ and\ \bibinfo {author} {\bibfnamefont {C.~P.}\
  \bibnamefont {Massen}},\ }\href {\doibase 10.1103/PhysRevE.71.016128}
  {\bibfield  {journal} {\bibinfo  {journal} {Phys. Rev. E}\ }\textbf {\bibinfo
  {volume} {71}},\ \bibinfo {pages} {016128} (\bibinfo {year}
  {2005})}\BibitemShut {NoStop}%
\bibitem [{\citenamefont {Varrato}\ and\ \citenamefont
  {Foffi}(2011)}]{Varrato2011}%
  \BibitemOpen
  \bibfield  {author} {\bibinfo {author} {\bibfnamefont {F.}~\bibnamefont
  {Varrato}}\ and\ \bibinfo {author} {\bibfnamefont {G.}~\bibnamefont
  {Foffi}},\ }\href {\doibase 10.1080/00268976.2011.640039} {\bibfield
  {journal} {\bibinfo  {journal} {Mol. Phys.}\ }\textbf {\bibinfo {volume}
  {109}},\ \bibinfo {pages} {2923} (\bibinfo {year} {2011})}\BibitemShut
  {NoStop}%
\bibitem [{\citenamefont {Kranz}\ \emph {et~al.}(2015)\citenamefont {Kranz},
  \citenamefont {Ara{\'{u}}jo}, \citenamefont {Andrade},\ and\ \citenamefont
  {Herrmann}}]{Kranz2015}%
  \BibitemOpen
  \bibfield  {author} {\bibinfo {author} {\bibfnamefont {J.~J.}\ \bibnamefont
  {Kranz}}, \bibinfo {author} {\bibfnamefont {N.~A.~M.}\ \bibnamefont
  {Ara{\'{u}}jo}}, \bibinfo {author} {\bibfnamefont {J.~S.}\ \bibnamefont
  {Andrade}}, \ and\ \bibinfo {author} {\bibfnamefont {H.~J.}\ \bibnamefont
  {Herrmann}},\ }\href {\doibase 10.1103/PhysRevE.92.012802} {\bibfield
  {journal} {\bibinfo  {journal} {Phys. Rev. E - Stat. Nonlinear, Soft Matter
  Phys.}\ }\textbf {\bibinfo {volume} {92}},\ \bibinfo {pages} {012802}
  (\bibinfo {year} {2015})}\BibitemShut {NoStop}%
\bibitem [{\citenamefont {Herrmann}\ \emph {et~al.}(1990)\citenamefont
  {Herrmann}, \citenamefont {Mantica},\ and\ \citenamefont
  {Bessis}}]{Herrmann1990}%
  \BibitemOpen
  \bibfield  {author} {\bibinfo {author} {\bibfnamefont {H.~J.}\ \bibnamefont
  {Herrmann}}, \bibinfo {author} {\bibfnamefont {G.}~\bibnamefont {Mantica}}, \
  and\ \bibinfo {author} {\bibfnamefont {D.}~\bibnamefont {Bessis}},\ }\href
  {http://journals.aps.org/prl/abstract/10.1103/PhysRevLett.65.3223} {\bibfield
   {journal} {\bibinfo  {journal} {Phys. Rev. Lett.}\ }\textbf {\bibinfo
  {volume} {65}},\ \bibinfo {pages} {3223} (\bibinfo {year}
  {1990})}\BibitemShut {NoStop}%
\bibitem [{\citenamefont {Baram}\ \emph {et~al.}(2004)\citenamefont {Baram},
  \citenamefont {Herrmann},\ and\ \citenamefont {Rivier}}]{Baram2004}%
  \BibitemOpen
  \bibfield  {author} {\bibinfo {author} {\bibfnamefont {R.~M.}\ \bibnamefont
  {Baram}}, \bibinfo {author} {\bibfnamefont {H.~J.}\ \bibnamefont {Herrmann}},
  \ and\ \bibinfo {author} {\bibfnamefont {N.}~\bibnamefont {Rivier}},\ }\href
  {\doibase 10.1103/PhysRevLett.92.044301} {\bibfield  {journal} {\bibinfo
  {journal} {Phys. Rev. Lett.}\ }\textbf {\bibinfo {volume} {92}},\ \bibinfo
  {pages} {044301} (\bibinfo {year} {2004})}\BibitemShut {NoStop}%
\bibitem [{\citenamefont {Oron}\ and\ \citenamefont
  {Herrmann}(2000)}]{Oron2000}%
  \BibitemOpen
  \bibfield  {author} {\bibinfo {author} {\bibfnamefont {G.}~\bibnamefont
  {Oron}}\ and\ \bibinfo {author} {\bibfnamefont {H.}~\bibnamefont
  {Herrmann}},\ }\href {\doibase 10.1088/0305-4470/33/7/310} {\bibfield
  {journal} {\bibinfo  {journal} {J. Phys. A Math. Gen.}\ }\textbf {\bibinfo
  {volume} {33}},\ \bibinfo {pages} {1417} (\bibinfo {year}
  {2000})}\BibitemShut {NoStop}%
\bibitem [{\citenamefont {Baram}\ and\ \citenamefont
  {Herrmann}(2005)}]{Baram2005}%
  \BibitemOpen
  \bibfield  {author} {\bibinfo {author} {\bibfnamefont {R.~M.}\ \bibnamefont
  {Baram}}\ and\ \bibinfo {author} {\bibfnamefont {H.~J.}\ \bibnamefont
  {Herrmann}},\ }\href {\doibase 10.1103/PhysRevLett.95.224303} {\bibfield
  {journal} {\bibinfo  {journal} {Phys. Rev. Lett.}\ }\textbf {\bibinfo
  {volume} {95}},\ \bibinfo {pages} {224303} (\bibinfo {year}
  {2005})}\BibitemShut {NoStop}%
\bibitem [{\citenamefont {{\AA}str{\"{o}}m}\ \emph {et~al.}(2000)\citenamefont
  {{\AA}str{\"{o}}m}, \citenamefont {Herrmann},\ and\ \citenamefont
  {Timonen}}]{Astrom2000}%
  \BibitemOpen
  \bibfield  {author} {\bibinfo {author} {\bibfnamefont {J.~A.}\ \bibnamefont
  {{\AA}str{\"{o}}m}}, \bibinfo {author} {\bibfnamefont {H.~J.}\ \bibnamefont
  {Herrmann}}, \ and\ \bibinfo {author} {\bibfnamefont {J.}~\bibnamefont
  {Timonen}},\ }\href {\doibase 10.1103/PhysRevLett.84.638} {\bibfield
  {journal} {\bibinfo  {journal} {Phys. Rev. Lett.}\ }\textbf {\bibinfo
  {volume} {84}},\ \bibinfo {pages} {638} (\bibinfo {year} {2000})}\BibitemShut
  {NoStop}%
\bibitem [{\citenamefont {{\AA}str{\"{o}}m}\ and\ \citenamefont
  {Timonen}(2012)}]{Astroma2012}%
  \BibitemOpen
  \bibfield  {author} {\bibinfo {author} {\bibfnamefont {J.~A.}\ \bibnamefont
  {{\AA}str{\"{o}}m}}\ and\ \bibinfo {author} {\bibfnamefont {J.}~\bibnamefont
  {Timonen}},\ }\href@noop {} {\bibfield  {journal} {\bibinfo  {journal} {Eur.
  Phys. J. E}\ }\textbf {\bibinfo {volume} {35}} (\bibinfo {year}
  {2012})}\BibitemShut {NoStop}%
\bibitem [{\citenamefont {McCann}\ \emph {et~al.}(1979)\citenamefont {McCann},
  \citenamefont {Nishenko}, \citenamefont {Sykes},\ and\ \citenamefont
  {Krause}}]{McCann1979}%
  \BibitemOpen
  \bibfield  {author} {\bibinfo {author} {\bibfnamefont {W.~R.}\ \bibnamefont
  {McCann}}, \bibinfo {author} {\bibfnamefont {S.~P.}\ \bibnamefont
  {Nishenko}}, \bibinfo {author} {\bibfnamefont {L.~R.}\ \bibnamefont {Sykes}},
  \ and\ \bibinfo {author} {\bibfnamefont {J.}~\bibnamefont {Krause}},\ }\href
  {\doibase 10.1007/BF00876211} {\bibfield  {journal} {\bibinfo  {journal}
  {Pure Appl. Geophys.}\ }\textbf {\bibinfo {volume} {117}},\ \bibinfo {pages}
  {1082} (\bibinfo {year} {1979})}\BibitemShut {NoStop}%
\bibitem [{\citenamefont {Lomnitz}(1982)}]{Society1982}%
  \BibitemOpen
  \bibfield  {author} {\bibinfo {author} {\bibfnamefont {C.}~\bibnamefont
  {Lomnitz}},\ }\href@noop {} {\bibfield  {journal} {\bibinfo  {journal} {Bull.
  Seism. Soc. Am.}\ }\textbf {\bibinfo {volume} {72}},\ \bibinfo {pages} {1411}
  (\bibinfo {year} {1982})}\BibitemShut {NoStop}%
\bibitem [{\citenamefont {Parker}(1995)}]{Parker1995}%
  \BibitemOpen
  \bibfield  {author} {\bibinfo {author} {\bibfnamefont {J.~R.}\ \bibnamefont
  {Parker}},\ }\href {\doibase 10.1016/0040-9383(94)00049-Q} {\bibfield
  {journal} {\bibinfo  {journal} {Topology}\ }\textbf {\bibinfo {volume}
  {34}},\ \bibinfo {pages} {489} (\bibinfo {year} {1995})}\BibitemShut
  {NoStop}%
\bibitem [{\citenamefont {{Kausch-Blecken von Schmeling}}\ and\ \citenamefont
  {Tschoegl}(1970)}]{Kausch1970}%
  \BibitemOpen
  \bibfield  {author} {\bibinfo {author} {\bibfnamefont {H.~H.}\ \bibnamefont
  {{Kausch-Blecken von Schmeling}}}\ and\ \bibinfo {author} {\bibfnamefont
  {N.~W.}\ \bibnamefont {Tschoegl}},\ }\href@noop {} {\bibfield  {journal}
  {\bibinfo  {journal} {Nature}\ }\textbf {\bibinfo {volume} {225}},\ \bibinfo
  {pages} {1119} (\bibinfo {year} {1970})}\BibitemShut {NoStop}%
\bibitem [{\citenamefont {Pickover}(1989)}]{Pickover1989}%
  \BibitemOpen
  \bibfield  {author} {\bibinfo {author} {\bibfnamefont {C.~A.}\ \bibnamefont
  {Pickover}},\ }\href@noop {} {\bibfield  {journal} {\bibinfo  {journal}
  {Comput. Graph.}\ }\textbf {\bibinfo {volume} {13}},\ \bibinfo {pages} {63}
  (\bibinfo {year} {1989})}\BibitemShut {NoStop}%
\bibitem [{\citenamefont {Baram}\ and\ \citenamefont
  {Herrmann}(2004)}]{Baram2004a}%
  \BibitemOpen
  \bibfield  {author} {\bibinfo {author} {\bibfnamefont {R.~M.}\ \bibnamefont
  {Baram}}\ and\ \bibinfo {author} {\bibfnamefont {H.}~\bibnamefont
  {Herrmann}},\ }\href {\doibase 10.1142/S0218348X04002549} {\bibfield
  {journal} {\bibinfo  {journal} {Fractals}\ }\textbf {\bibinfo {volume}
  {12}},\ \bibinfo {pages} {293} (\bibinfo {year} {2004})}\BibitemShut
  {NoStop}%
\bibitem [{\citenamefont {Baram}(2005)}]{Baram2005a}%
  \BibitemOpen
  \bibfield  {author} {\bibinfo {author} {\bibfnamefont {R.~M.}\ \bibnamefont
  {Baram}},\ }\emph {\bibinfo {title} {{Polydisperse granular packings and
  bearings}}},\ \href {http://www.ciul.ul.pt/{~}reza/{\_}Media/thesis.pdf}
  {Ph.D. thesis} (\bibinfo {year} {2005})\BibitemShut {NoStop}%
\bibitem [{\citenamefont {St{\"{a}}ger}\ \emph {et~al.}(2016)\citenamefont
  {St{\"{a}}ger}, \citenamefont {Ara{\'{u}}jo},\ and\ \citenamefont
  {Herrmann}}]{Stager2016}%
  \BibitemOpen
  \bibfield  {author} {\bibinfo {author} {\bibfnamefont {D.~V.}\ \bibnamefont
  {St{\"{a}}ger}}, \bibinfo {author} {\bibfnamefont {N.~A.~M.}\ \bibnamefont
  {Ara{\'{u}}jo}}, \ and\ \bibinfo {author} {\bibfnamefont {H.~J.}\
  \bibnamefont {Herrmann}},\ }\href {\doibase 10.1103/PhysRevLett.116.254301}
  {\bibfield  {journal} {\bibinfo  {journal} {Phys. Rev. Lett.}\ }\textbf
  {\bibinfo {volume} {116}},\ \bibinfo {pages} {254301} (\bibinfo {year}
  {2016})}\BibitemShut {NoStop}%
\end{thebibliography}%

\appendix
\section{Mathematics of circle inversion}\label{app:math_of_inversion}

Circle inversion is simplified by using inversion coordinates for circles and inversion circles. We generalize in the following to any dimension higher or equal to two. Therefore, we refer to $n$-spheres with $n \geq 1$, where a circle and a sphere are a $1$-sphere and a $2$-sphere, respectively. The inversion coordinates $(a_1,a_2,\ldots,a_{n+3})$ of an $n$-sphere are defined by its center $\vec{x}=(x_1,\ldots,x_{n+1})$ and its radius $r$ as
\begin{equation}
\begin{aligned}
a_i&=\frac{x_i}{r},\  \text{for} \ i=\{1,\ldots,n+1\} \\ a_{n+2}&=\frac{x_1^2+\ldots + x_{n+1}^2 -r^2-1}{2r}, \\ a_{n+3}&=\frac{x_1^2+\ldots + x_{n+1}^2 -r^2+1}{2r}.
\end{aligned}\label{eq:inversioncoordinates}
\end{equation}
They satisfy the relation $a_1^2+\ldots+a_{n+2}^2-a_{n+3}^2=1$ and one can express the usual parameters as
\begin{equation}
\begin{aligned}
r&=\frac{1}{a_{n+3}-a_{n+2}}  , \\
x_i&=\frac{a_i}{a_{n+3}-a_{n+2}}, \ \text{for} \ i=\{1,\ldots,n+1\}. \\
\end{aligned}\label{eq:realcoordinates}
\end{equation}
By using the inversion coordinates for an $n$-sphere, the inversion of a sphere (with coordinates $a_i$) leads to the image ($a_i'$) by
\begin{equation}
(a_1',\ldots,a_{n+3}')^\top = \mathbf{M} \cdot (a_1,\ldots,a_{n+3})^\top,
\end{equation}
where the superscript $\top$ denotes the transpose, and the $(n+3) \times (n+3)$ matrix $\mathbf{M}$ is defined by the coordinates $A_i$ of the inversion sphere at which we invert at as
\begin{equation}
\mathbf{M}=\mathbf{I}-2(A_1, \ldots, A_{n+3})^\top \cdot (A_1, \ldots, A_{n+2}, -A_{n+3}),\label{eq:inversion}
\end{equation}
where $\mathbf{I}$ is the identity matrix. A more detailed derivation of the matrix $\mathbf{M}$ can be found in Ref.\ \cite{Baram2005a}. Note that we consider a sphere with $r>0$ as filled, referring to the space inside the sphere, and a sphere with radius $r<0$ as a hole, referring to the space outside the sphere of radius $|r|$. In the special case of $a_{n+2}=a_{n+3}$, we have a half-space defined by the plane with normal vector $(a_1,...,a_{n+1})$ and distance $a_{n+2}$ from the origin, covering the space in the direction of the normal vector.

\section{Solving for the positions and radii of the setup elements}\label{app:solve_for_spatial_details}

Apart from simple trigonometry we will in many calculations use the relation for two intersecting spheres, which is valid in any dimension, that states
\begin{equation}
\label{eq:intersecting_spheres}
d^2=r_1^2+r_2^2 + 2r_1 r_2 \cos \alpha,
\end{equation}
where $d$ is the distance between the centers of two spheres with radii $r_1$ and $r_2$ that intersect with angle $\alpha$. For spheres perpendicular to each other Eq.\ (\ref{eq:intersecting_spheres}) simplifies to 
\begin{equation}
\label{eq:intersecting_spheres_perpendicular}
d^2=r_1^2+r_2^2.
\end{equation}

\begin{figure}[]
\begin{center}
	\includegraphics[width=.5\columnwidth]{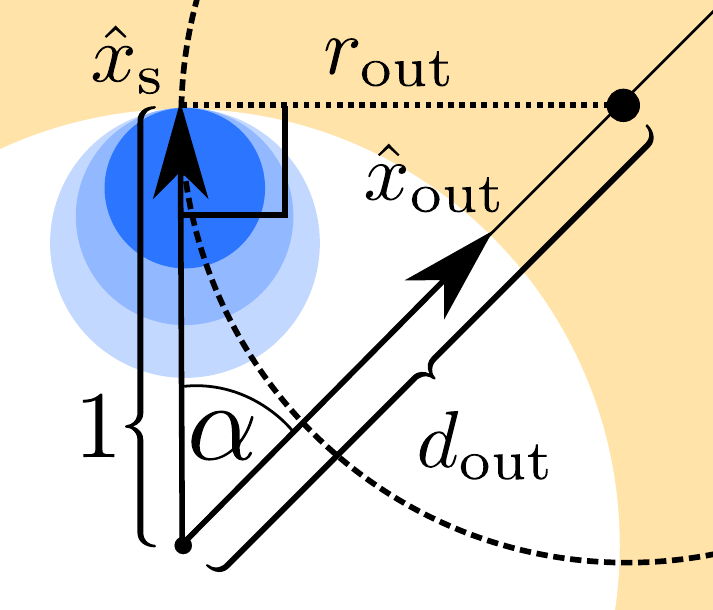}
\end{center}
\caption{
\label{fig:I1_calc} Solving for the details of the outer inversion spheres: In the right triangle shown, we find that the distance $d_\text{out}=1/\cos\alpha=1/(\hat{x}_\text{out} \cdot \hat{x}_{s})$ and the radius $r_\text{out}=\sqrt{d_\text{out}^2-1}$
}
\end{figure}

Every element of the setup is defined by the unit vector of its position, the distance of its center to the center of the unit sphere, and its radius. By choosing a regular polygon on which we base our generating setup, we define the unit vectors of the outer inversion spheres and the primary seeds. The distance of the outer inversion spheres $d_\text{out}$ and their radius $r_\text{out}$ is defined, given the fact that they need to be perpendicular to the unit sphere and the nearest primary seeds. In the right triangle shown in Fig.\ \ref{fig:I1_calc}, we find $d_\text{out}=1/\cos\alpha=1/(\hat{x}_\text{out} \cdot \hat{x}_{s})$, where $\hat{x}_\text{out}$ and $\hat{x}_{s}$ are the unit vectors of an outer inversion sphere and a nearest neighboring primary seed, respectively. Because the outer inversion sphere is perpendicular to the unit sphere, we find $r_\text{out}=\sqrt{d_\text{out}^2-1}$.

Next we want to solve for the inner inversion spheres, which are different for the two families F1 and F2.

\begin{figure}[]
\begin{center}
	\includegraphics[width=\columnwidth]{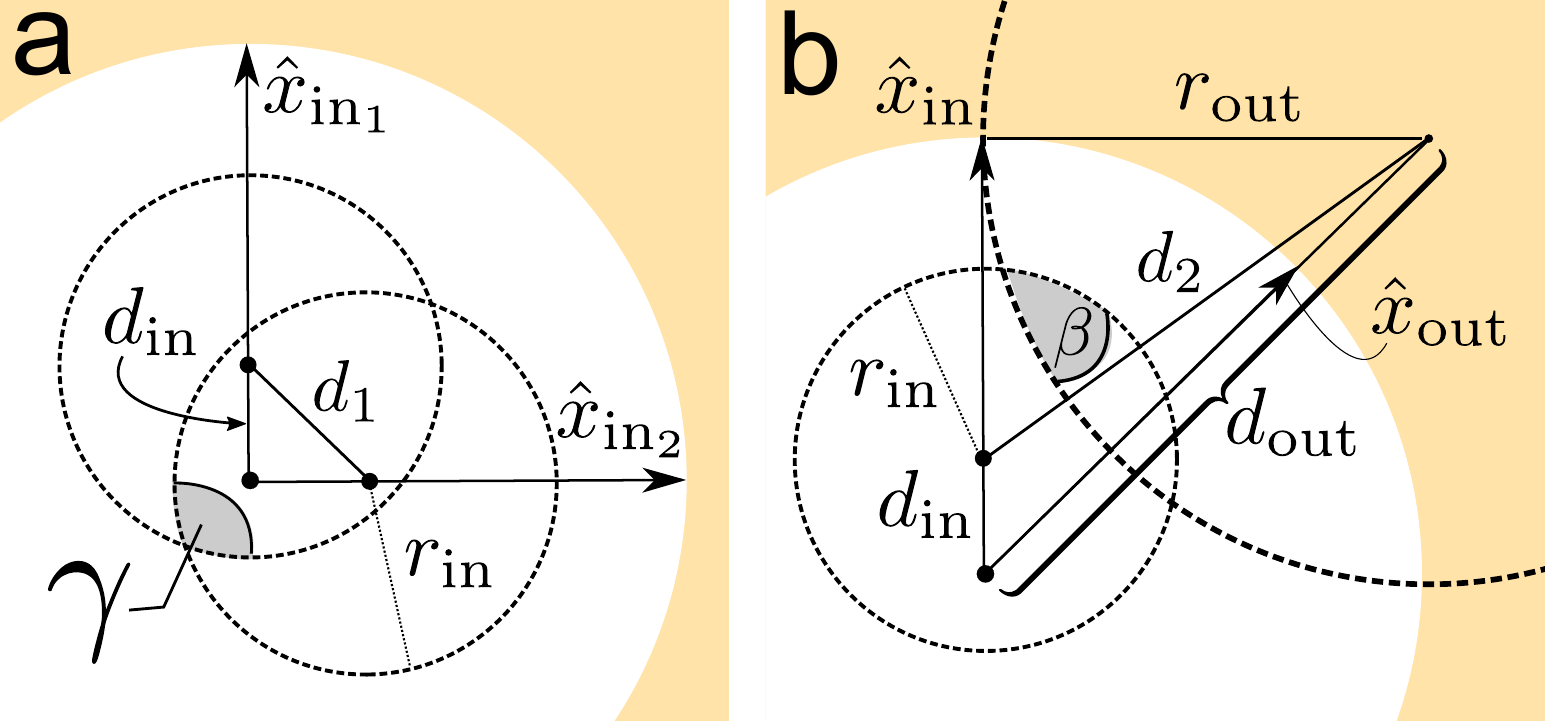}
\end{center}
\caption{
\label{fig:I2_calc_F1} Solving for the details of the inner inversion spheres for F1: In (a) we find for $d_1$ that $d_1^2 = 2 r_\text{in}^2 (1+\cos \gamma)=d_\text{in}^2(1-\hat{x}_{\text{in}_1} \cdot \hat{x}_{\text{in}_2})$. In (b) we find for $d_2$ that $d_2^2= d^2_\text{out} + d^2_\text{in} - 2 d_\text{out}d_\text{in} \hat{x}_\text{in} \cdot \hat{x}_\text{out} = r^2_\text{out} + r^2_\text{in} + 2r^2_\text{out}r^2_\text{in} \cos \beta$.
}
\end{figure}

For F1, the unit vectors of the inner inversion spheres are identical to the ones of the primary seeds. One can express the distance $d_1$ between two nearest inner inversion spheres, as shown in Fig.\ \ref{fig:I2_calc_F1}a, in two different ways to get the equation

\begin{equation}
\label{eq:F1gamma}
(d_1^2=) \ \ 2 r_\text{in}^2 (1+\cos \gamma)=d_\text{in}^2(1-\hat{x}_{\text{in}_1} \cdot \hat{x}_{\text{in}_2}),
\end{equation}
where $\hat{x}_{\text{in}_1}$ and $\hat{x}_{\text{in}_2}$ are the unit vectors of two nearest inner inversion spheres. From the fact that an inner inversion sphere intersects its closest outer inversion spheres with angle $\beta$, one can express the distance $d_2$ between them shown in Fig.\ \ref{fig:I2_calc_F1}b in two different ways to get the equation
\begin{multline}
\label{eq:F1beta}
(d_2^2=) \ \ d^2_\text{out} + d^2_\text{in} - 2 d_\text{out}d_\text{in} \hat{x}_\text{in} \cdot \hat{x}_\text{out} \\ = r^2_\text{out} + r^2_\text{in} + 2r^2_\text{out}r^2_\text{in} \cos \beta,
\end{multline}
where $\hat{x}_{\text{in}}$ and $\hat{x}_{\text{out}}$ are the unit vectors of a closest pair of inner and outer inversion spheres. One can find $r_\text{in}$ and $d_\text{in}$ from Eqs.\ (\ref{eq:F1gamma}) and (\ref{eq:F1beta}). 

\begin{figure}[]
\begin{center}
	\includegraphics[width=\columnwidth]{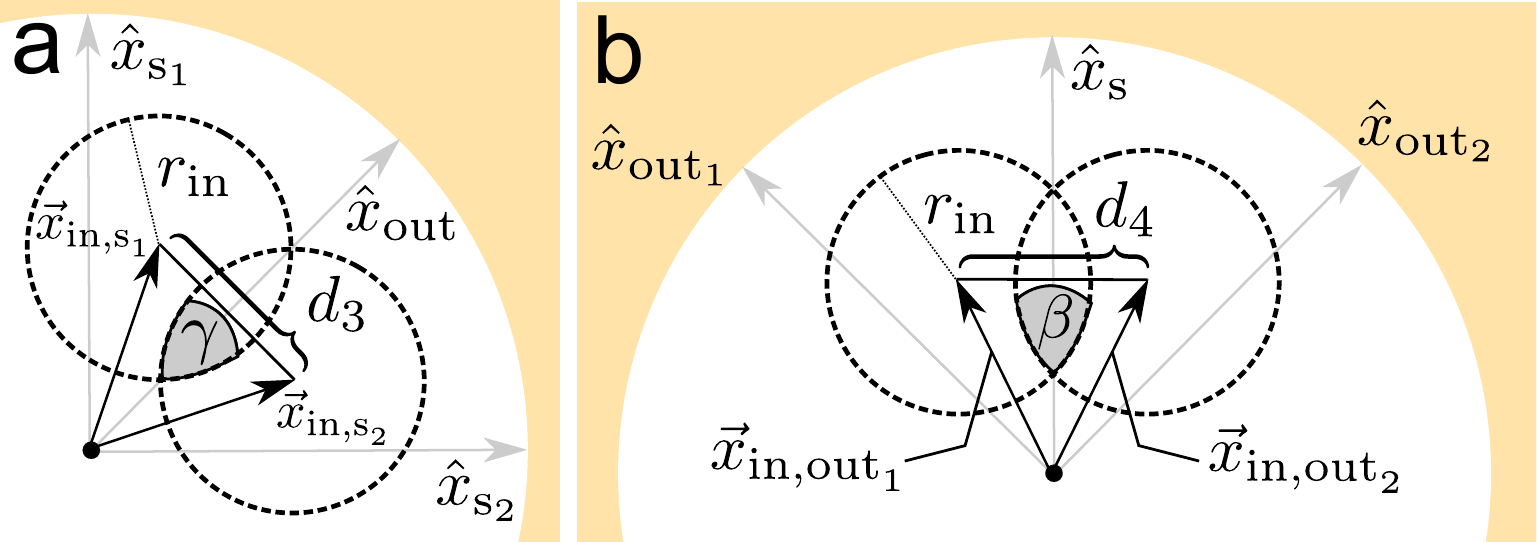}
\end{center}
\caption{
\label{fig:I2_calc_F2} Solving for the details of the inner inversion spheres for F2: (a) For $d_3$ we find $d_3^2 = (\vec{x}_{\text{in,s}_1}-\vec{x}_{\text{in,s}_2})^2=2 r_\text{in}^2 (1+\cos \gamma)$. (b) For $d_4$ we find  $d_4^2= (\vec{x}_{\text{in,out}_1}-\vec{x}_{\text{in,out}_2})^2=2 r_\text{in}^2 (1+\cos \beta)$.
}
\end{figure}

For F2, the unit vector of an inner inversion sphere $\hat{x}_\text{in}$ is a combination of the one of its nearest outer inversion sphere $\hat{x}_\text{out}$ and the one of its nearest primary seed $\hat{x}_\text{s}$ such that we can write 
\begin{equation}
\label{eq:F2unitvector}
\hat{x}_\text{in} = p \hat{x}_\text{out} + q \hat{x}_\text{s}. 
\end{equation}
The condition for this vector to be a unit vector gives us the equation
\begin{equation}
\label{eq:F2unitvector_condition}
(\hat{x}_\text{in}^2) = p^2 + q^2 + 2pq\hat{x}_\text{out} \cdot \hat{x}_\text{s} =1.
\end{equation}
Every inner inversion sphere is perpendicular to the nearest outer inversion sphere such that 
\begin{equation}
\label{eq:F2_inner_perp_outer1}
(d_\text{out} \hat{x}_\text{out}- d_\text{in} \hat{x}_\text{in})^2 = r_\text{out}^2+r_\text{in}^2,
\end{equation}
where $\hat{x}_{\text{in}}$ and $\hat{x}_{\text{out}}$ are the unit vectors of a closest pair of inner and outer inversion spheres. Using Eq.\ (\ref{eq:F2unitvector}), this gives us 
\begin{equation}
\label{eq:F2_inner_perp_outer2}
d_\text{out}^2 + d_\text{in}^2 -2d_\text{out}d_\text{in}(p+q \hat{x}_\text{out} \cdot  \hat{x}_\text{in}) = r_\text{out}^2+r_\text{in}^2.
\end{equation}
From the fact that two inner inversion spheres that are nearest neighbors of an outer inversion sphere intersect with angle $\gamma$, we can express the distance $d_3$ shown in Fig.\ \ref{fig:I2_calc_F2}a in two different ways as
\begin{equation}
\label{eq:F2gamma}
(d_3^2=)\ \ (\vec{x}_{\text{in,s}_1}-\vec{x}_{\text{in,s}_2})^2=2 r_\text{in}^2 (1+\cos \gamma),
\end{equation}
with $\vec{x}_{\text{in,s}_1}$ and $\vec{x}_{\text{in,s}_2}$ being the unit vectors of two inner inversion spheres that are closest to an outer inversion sphere. Since $\vec{x}_{\text{in,s}_1}=d_\text{in} (p\hat{x}_\text{out} + q \hat{x}_{\text{s}_1})$ and $\vec{x}_{\text{in,s}_2}=d_\text{in} (p\hat{x}_\text{out} + q \hat{x}_{\text{s}_2})$, where $\hat{x}_{\text{s}_1}$ and $\hat{x}_{\text{s}_2}$ are the unit vectors of two primary seeds that are nearest to an outer inversion sphere with the unit vector $\hat{x}_{\text{out}}$, we find from Eq.\ (\ref{eq:F2gamma}) that
\begin{equation}
\label{eq:F2gamma2}
d_\text{in}^2 q^2 (2-2\hat{x}_{\text{s}_1} \cdot \hat{x}_{\text{s}_2})=2 r_\text{in}^2 (1+\cos \gamma).
\end{equation}
In analogy to the derivation of Eq.\ (\ref{eq:F2gamma2}), we can derive a similar equation for the angle $\beta$ by expressing the the distance $d_4$ as shown in Fig.\ \ref{fig:I2_calc_F2}b in two different ways to finally find
\begin{equation}
\label{eq:F2beta2}
d_\text{in}^2 p^2 (2-2\hat{x}_{\text{out}_1} \cdot \hat{x}_{\text{out}_2})=2 r_\text{in}^2 (1+\cos \beta).
\end{equation}
One can now find $p$, $q$, $r_\text{in}$, and $d_\text{in}$ from the four Eqs.\ (\ref{eq:F2unitvector_condition}), (\ref{eq:F2_inner_perp_outer2}), (\ref{eq:F2gamma2}), and (\ref{eq:F2beta2}). The unit vector $\hat{x}_{\text{in}}$ can then be found from Eq.\ (\ref{eq:F2unitvector}).

After solving for the inner inversion spheres, we can find the radius for the primary seeds $r_\text{s}$. For both F1 and F2, we get from the fact that every primary seed is perpendicular to a closest inner inversion sphere the equation
\begin{equation}
\label{eq:Rseed}
r_\text{s}^2+r_\text{in}^2=(d_\text{s} \hat{x}_\text{s} - d_\text{in} \hat{x}_\text{in})^2=d_\text{s}^2+d_\text{in}^2-2d_\text{s}d_\text{in} \hat{x}_\text{s} \cdot \hat{x}_\text{in},
\end{equation}
where $d_\text{s}$ is the distance from the center of a primary seed to the center of the unit sphere. Since the primary seeds are tangent to the unit sphere, we have $d_\text{s}=1-r_\text{s}$, which together with Eq.\ (\ref{eq:Rseed}) can be solved for $d_\text{s}$ and $r_\text{s}$.

Finally there might be uncovered space that can be filled by additional seeds. An additional seed in the center of the packing is needed if the inner inversion spheres do not cover it, i.e., for $r_\text{in} < d_\text{in}$. Since this seed would need to be perpendicular to the inner inversion spheres, its radius is defined as $r_\text{center}=\sqrt{d_\text{in}^2-r_\text{in}^2}$. Further additional seeds might be needed between inner and outer inversions spheres. They can only lay in the directions of the edges, faces, etc., of the convex polytope whose vertices are at the positions of the primary seeds. For a given unit vector $\hat{x}_{\text{as}}$ of the position $\vec{x}_{\text{as}} = d_{\text{as}} \hat{x}_{\text{as}}$ of such an additional seed, one can find its distance $d_{\text{as}}$ and its radius $r_{\text{as}}$ from the fact that the seed would need to be perpendicular to the closest inner and outer inversion spheres. From the fact that the seed is perpendicular to a closest inner inversion sphere at position $\vec{x}_\text{in}=d_\text{in} \hat{x}_\text{in}$ and a closest outer inversion sphere at position $\vec{x}_\text{out} = d_\text{out} \hat{x}_\text{out}$, we find
\begin{align}\label{eq:addseed1}
r_{\text{as}}^2+r_{\text{in}}^2 &= (\vec{x}_{\text{as}}-\vec{x}_\text{in})^2 = d_{\text{as}}^2 + d_\text{in}^2 - d_{\text{as}}d_\text{in}\hat{x}_{\text{as}} \cdot \hat{x}_\text{in},\\ \label{eq:addseed2}
r_{\text{as}}^2+r_{\text{out}}^2 &= (\vec{x}_{\text{as}}-\vec{x}_\text{out})^2 = d_{\text{as}}^2 + d_\text{out}^2 - d_{\text{as}}d_\text{out}\hat{x}_{\text{as}} \cdot \hat{x}_\text{out}.
\end{align}
The system of equations (\ref{eq:addseed1}) and (\ref{eq:addseed2}) has a unique solution for $d_{\text{as}}>0$ and $r_{\text{as}}>0$, if there is uncovered space between the inner and outer inversion spheres along $\hat{x}_{\text{as}}$.

\section{Generate packings computationally efficiently}\label{app:computation}
Starting with the seeds, one can invert every of them at every inversion sphere of the generating setup. One can iteratively repeat that procedure with all newly generated spheres. But one needs to take care of not producing any sphere that is already present. If one inverts a sphere at an inversion sphere that is perpendicular to the sphere, the generated sphere is identical to the original sphere. Additionally, different sequences of inversions can lead to the same sphere.

Fortunately, there is a simple trick to avoid generating any sphere twice which we adopted from Ref.\ \cite{Borkovec1994}. One can ensure to generate each sphere only once by only inverting a sphere at an inversion sphere if the inverse has its center inside the \emph{target region} of the corresponding inversion sphere. The target regions of the inversion spheres are such that they are regions inside the corresponding inversion spheres such that no two target regions overlap but their union equals the union of the all inversion spheres. A simple way to define target regions for a set of ordered inversion spheres is to define the target region of the first inversion sphere as the whole inside of it, and the target region of every following inversion sphere as the inside of the inversion sphere minus the overlap with previous inversion spheres. Numerically, one can neglect the surface of the inversion spheres, since only infinitely small spheres could end up having their center on the surface of an inversion sphere, given that spheres can only be perpendicular to inversion spheres.

With this method, the generation of all spheres larger than a certain smallest radius becomes a simple branching process where every new inverse of a sphere is smaller than the original, so one can cut a branch if all possible inversions lead to spheres that are smaller than the considered smallest radius. This allows to calculate the functions to estimate the fractal dimensions as described in Sec.\ \ref{sec:df} computationally using very few memory.

\end{document}